\documentstyle[11pt,epsf,rotate,aaspp4]{article}

\newcommand{\be}{\begin{equation}}
\newcommand{\ee}{\end{equation}}
\newcommand{\ben}{\begin{eqnarray}}
\newcommand{\een}{\end{eqnarray}}
\newcommand{\pl}{\partial}

\received{}
\revised{}
\accepted{September 30, 1997}
\cpright{}{}

\journalid{}{}
\articleid{}{}
\paperid{}

\cpright{}{}
\ccc{}

\slugcomment{}

\lefthead{Papaloizou, Terquem \& Lin}
\righthead{ Self--gravitating near Keplerian warped disks}
\begin{document}

\title{On the global warping of a thin self--gravitating near
Keplerian gaseous disk with application to the disk in NGC~4258}
 
\vspace*{0.5cm}

\author{John C. B. Papaloizou\altaffilmark{1}, Caroline
Terquem\altaffilmark{2,1,3} and Doug N. C. Lin\altaffilmark{2}}

\altaffiltext{1}{Astronomy Unit, School of Mathematical Sciences,
Queen Mary~\& Westfield College, University of London, Mile End Road,
London E1 4NS, UK -- J.C.B.Papaloizou@qmw.ac.uk}

\altaffiltext{2}{UCO/Lick Observatory, University of California,
Santa--Cruz, CA~95064, USA -- ct,~lin@ucolick.org}

\altaffiltext{3}{Laboratoire d'Astrophysique, Universit\'e
J. Fourier/CNRS, BP 53, 38041 Grenoble Cedex~9, France}

\vspace*{2.cm}

{\centerline{ To appear in ApJ}}

\newpage

\begin{abstract}

\noindent We derive the tilt equation governing the inclination of a
thin self--gravitating gaseous disk subject to low frequency global
$m=1$ bending perturbations. The disk orbits under the influence of a
dominant central mass. However, self--gravity can be important enough
that the disk approaches marginal stability to local axisymmetric
perturbations ($Q\sim 1$). The vertical restoring forces due to
self--gravity and pressure are evaluated correct to first order in the
aspect ratio $H/r.$ Thus the effects of bending waves are included
correct to lowest order in the wave crossing rate $(H/r)\Omega, $
$\Omega$ being a characteristic disk rotation frequency.

\noindent Both free and forced disturbances are considered. The disk
response and precession frequency induced by the presence of a binary
companion in an orbit with general inclination to the unperturbed disk
plane are derived. When the degree of warping and the inclination are
small, it is shown that identical results are obtained if,
alternatively, perturbation of the disk out of an equilibrium plane
coinciding with that of the companion is considered, the time averaged
potential due to the latter being incorporated into the equilibrium
potential. The condition for the disk to precess approximately like a
rigid body with a small degree of warping is found to be that the
density wave crossing time be significantly shorter than the
precession period. We consider the effects of the presence of a
viscosity which can be characterized with the standard $\alpha$
parameterization and find that, to the order we work, the precession
frequency is unaffected, with any change to the inclination of the
slightly warped disk occuring at a slower rate. For $\alpha \ll H/r,$
effects due to both pressure and self--gravity are important in the
response, while for $\alpha \gg H/r,$ the response becomes dominated
by self--gravity with pressure effects becoming negligible.
 
\noindent As an application of these results, we explore the
possibility that the recently observed warped disk in the active
galaxy NGC~4258 is produced by a binary companion. Our results
indicate that it may be produced by a companion with a mass comparable
to or larger than that of the observed disk. If the only source of
warping is such a companion, a small trailing twist is produced by
viscosity for modest $\alpha \sim 0.1.$ The outer edge of the disk may
also be tidally truncated by the companion.

\end{abstract}


\newpage

\section{Introduction}

\label{sec:intro}

\noindent Gaseous disk configurations which are dominated by a central
mass and in which internal self--gravity is important may occur in
different astrophysical contexts such as protostellar disks
(Larson~1984) or active galactic nuclei (Paczynski~1977).

\noindent The most definitive evidence for a gaseous accretion disk in
active galactic nuclei is provided by the discovery of megamasers
(Claussen, Heiligman~\& Lo~1984) around the nucleus of the mildly
active galaxy NGC~4258 (Claussen~\& Lo~1986). Confirmation of
Keplerian rotation was obtained from the radio interferometer (Nakai,
Inoue~\& Miyoshi~1993) and VLBI observations (Greenhill {\it et
al.}~1995). Based on the correlation between the spatial locations
and radial velocities of the masers, Miyoshi {\it et al.}~(1995)
deduced that the masers are located at $R \sim 0.13-0.26$~$pc$ around
a black hole with a mass $ 3.6 \times 10^7$~$M_\odot$ (Watson~\&
Wallin~1994; Maoz~1995). Recently the inner part of the disk has been
modelled with an advection--dominated flow (Lasota {\it et al.}~1996).

\noindent Random deviations from the Keplerian rotation curve at the
location of the masers are $\sim 3.5$~$km/s$. The lack of systematic
deviations provides an upper limit for the disk mass of $ 4\times
10^6$~$M_{\odot}$. The disk scale height $H< 3 \times 10^{-4}$~$pc$ is
such that $H/R < 2.5 \times 10^{-3}$ and the mid--plane temperature
$T_c < 10^3$~$K$ (Moran {\it et al.}~1995). In addition, the
high--velocity maser sources have negligible acceleration and are not
colinear with the low--velocity masers (Greenhill {\it et
al.}~1995). The radial dependence of declination of the red--shifted
(with respect to the systemic velocity of NGC~4258) high--velocity
masers is antisymmetric to that of the blue--shifted high--velocity
masers (Miyoshi {\it et al.}~1995). One scenario to account for these
observed properties is that the position angle of the rotation axis
varies continually with radius by up to 0.2 radians (Herrnstein,
Greenhill~\& Moran~1996). This corresponds to introducing a small warp
into the disk model. The inclination of the local orbital plane with
respect to the orbital plane at the innermost region of the disk, $g$,
being the ratio of the vertical displacement to the local radius,
varies with the disk radius. In at least one set of models,
small--amplitude variations in the angle between the line of sight and
the rotation axis are also introduced. This corresponds to introducing
a ``twist'' into the warped disk. The magnitude and direction of the
twist are not yet well constrained.

\noindent In this paper, we study the dynamics of the warped disk in
NGC~4258. The main constraints for any theoretical model are: 1) the
coherence and the smooth radial dependence in the inclination $g$, 2)
the negligible acceleration in the high velocity maser sources, and 3)
the nearly perfect Keplerian rotation curve.

\noindent We explore the possibility that the observed warp may be
produced by a binary companion which orbits in a plane inclined to
that of the disk.  Such non coplanar system could be produced if the
companion approaches the disk as a result of random gravitational
perturbations arising from the field stars in the nucleus of NGC~4258
(see \S~\ref{sec:discussion}). Indeed, in that case there is no {\it a
priori} reason to suppose alignment between its orbital angular
momentum vector and that of the matter which supplies the disk at
large distances. Furthermore, in the evolutionary situation we
envisage, the disk is not expected to have settled to an equilibrium
in which its mid--plane coincides with the companion's orbital plane
(see \S~\ref{sec:settle2}). Capture of a companion is not expected to
result systematically in the disruption of the disk. The orbit may
have indeed become bound when the companion was far away from the
disk, with a subsequent decrease of the separation due to dynamical
friction.

\noindent In the model we shall adopt in this paper, the disk is
subject to forces due to internal self--gravity, pressure, viscosity,
and also to the gravitational forces arising from the central object
and the perturbing companion on an inclined orbit.  Since we shall
focus on the secular response of the disk (see
\S~\ref{sec:potential}), the perturbation is the same as that produced
by a mass distributed uniformly along an inclined ring. For a near
Keplerian self--gravitating disk with small viscosity, pressure may in
principle be as important as self--gravity in determining the
response. However, for the model we adopt in \S~\ref{sec:NGC} for
NGC~4258, self--gravity turns out to dominate over pressure and
viscosity in determining the disk response to the perturbing
potential. Before going on to develop the tilt equation we use to
describe near Keplerian disks, we summarize some general results on
the dynamics of warped disks.

\subsection{Precession of warped disks}

\noindent In a classic paper, Hunter~\& Toomre~(1969) showed that an
isolated self--gravitating disk subject to a vertical displacement
generally precesses differentially, and thus cannot sustain a warped
configuration.  However, differential precession is prevented by
gravitational torques from the distorted disk itself if the disk has a
sharp edge. A clear physical description of how self--gravity acts to
smooth out differential precession is given by Toomre~(1983). A
similar process occurs if the disk orbits in the external potential
due to a companion (Hunter~\& Toomre~1969) or a flattened halo whose
equatorial plane is misaligned with the disk plane (Toomre~1983;
Dekel~\& Shlosman~1983; Sparke~1984; Sparke~\& Casertano~1988),
provided departure from spherical symmetry of the external potential
is relatively small. In this case, the disk settles into a discrete
bending mode (representing a warp) which is referred to as the
modified tilt mode because in the limit that the external potential is
spherically symmetric it reduces to the trivial rigid tilt mode. The
difference in shape and frequency between the rigid and modified tilt
modes is due to the fact that, in a non spherically symmetric
potential, the disk has to bend to alter the precession frequency at
each radius so that the rate is everywhere the same (Sparke~\&
Casertano~1988; Hofner~\& Sparke~ 1994). We note that the potential of
an oblate halo can be modeled by that of a distant massive fixed ring
which is misalinged with the disk plane (Toomre~1983). This situation
is analogous to that being investigated here in this paper, namely a
near Keplerian self--gravitating disk dominated by a central mass and
a smaller orbiting companion which in a time average sense gives the
appearance of a massive ring.

\noindent Papaloizou~\& Terquem~(1995) have shown that radial pressure
forces in a non self--gravitating inviscid accretion disk are also
able to smooth out differential precession. This process can be
effective if the sound crossing time through the disk is much smaller
than the precession period. When this condition is satisfied, bending
waves are able to propagate through the disk sufficiently fast so that
the different parts of the disk can ``communicate'' with each other
and adjust their precession rate to a constant value. This also
happens when viscosity is present, but the communication becomes
diffusive rather than wave--like when the Shakura~\& Sunyaev~(1973)
viscosity parameter $\alpha$ significantly exceeds the ratio of disk
semi--thickness to radius (see Papaloizou~\& Pringle~1983;
Demianski~\& Ivanov~1997). The theoretical expectation of uniform
precession of inclined disks under conditions of adequate physical
communication has been confirmed by the numerical simulations of
Larwood {\it et al.}~(1996).

\noindent When the disk precesses uniformly in an external potential,
its precession frequency can be calculated from the condition that the
disk is stationary in a rotating frame. This condition implies that
the net torque exerted by the external potential and the Coriolis
force is zero (by virtue of Newton's third law, the disk does not
exert a torque on itself). Kuijken~(1991) has used this condition to
calculate the precession frequency of a self--gravitating disk in a
flattened halo whose plane of symmetry is misaligned with the disk
plane. In the limit of small misalignment, his result reduces to that
given by perturbation theory (Sparke~\& Casertano~1988). As expected,
his results also indicate that when the departure from spherical
symmetry of the potential gets too large, self--gravity can no longer
maintain a uniform precession of the entire disk. In the context of
simulations of gaseous disks in interacting binary stars that were
misaligned with the binary orbital plane, Larwood {\it et al.}~(1996)
found that when the tidal potential of the companion is large enough,
forces due to pressure and viscosity in a non self--gravitating thin
disk cannot prevent differential precession.

\subsection{Timescale for settling to the warp mode}

\noindent We have pointed out above that the disk settles into a
(modified tilt) mode which enables it to undergo rigid
precession. This asymptotic state is possible because bending waves
transport away the energy associated with the transient response
(Toomre~1983). In a purely self--gravitating disk, this energy is
carried out towards the disk outer edge by bending waves whose group
velocity approaches zero there. The waves pile up with very short
wavelength at the edge where they would be expected to dissipate. The
timescale for settling to the warp mode is given by the characteristic
time for density waves to propagate through the disk but the process
is slowed down in the low surface density regions near the outer edge.
Hofner~\& Sparke~(1994) found that when a galactic disk is subject to
the potential of a misaligned flattened halo, the inner parts settle
first and the outer parts are unlikely to settle into the warp mode
within a Hubble time.

\noindent When pressure is present the waves may be reflected from the
edge before they attain arbitrary short wavelength there. Some
dissipation is then needed for the disk to settle into a state of near
rigid precession. This can be provided by disk viscosity which is
least effective on a global disturbance such as near rigid body
precession. Other transient disturbances would, as we have indicated
above, have shorter wavelengths and accordingly would be expected to
dissipate significantly more rapidly. In this context note that a warp
mode which deviates only slightly from a rigid tilt has a wavelength
which is much larger than the disk. Larwood {\it et al.}~(1996)
considered the disk response due to the perturbation of a
companion. In their simulations, of non self--gravitating inclined
disks with pressure and viscosity, the disk was seen to quickly settle
into a state of near rigid body precession. The time for this to
happen was consistent with about $\sim 1/\alpha$ orbital periods, this
being the decay time of short wavelength bending waves without
self--gravity (see Papaloizou~\& Lin~1994, 1995).

\noindent In this paper, we consider the limiting case of a very thin
self--gravitating disk (see \S~\ref{sec:parameters}) in which bending
waves propagate with a speed comparable to that of sound and the
settling timescale into the warp mode is short compared to the viscous
evolution timescale of the entire disk.

\subsection{Timescale for settling to the orbital plane}
\label{sec:settle2}

\noindent Papaloizou~\& Terquem~(1995) have shown that if a non
self--gravitating disk perturbed by a companion on an inclined orbit
evolves towards the symmetry plane of the potential (which does not
necessarily occur), it does so on a timescale related to the rate at
which angular momentum is transported to the companion which is
expected to be at least the disk viscous timescale. This conclusion
might be expected since the disk inclination as a whole can change
only if angular momentum is transferred between different parts of the
disk and then between the disk and the companion's orbit. As the long
term disk evolution is driven through the action of viscous stress,
adjustment should occur so that angular momentum transfer to the
orbit, and hence changes to disk inclination occur on that timescale
or longer. Such an adjustment can occur for example through evolution
of the location of the disk outer edge away from the companion's
orbit.

\noindent This was seen in the numerical simulations of Larwood {\it
et al.}~(1996) and Larwood~(1997), in which disk inclination was
ultimately seen to evolve much more slowly than the precession
rate. We also show that this is expected to occur for a mildly warped
self--gravitating disk in section \ref{sec:viscosity} below. Thus we
do not necessarily expect the inclination of the disk to have evolved
so that it coincides with the orbital plane.

\noindent Results of numerical simulations of the dynamical evolution
of rings highly inclined to the symmetry plane of some galactic
potential (Katz~\& Rix~1992; Christodoulou {\it et al.}~1992;
Christodoulou~\& Tohline~1993) are also in agreement with the above
conclusions. However, we emphasize that they only hold if the disk is
able to find a state in which differential precession can be
controlled by internal self--gravity and pressure. Settling toward a
preferred orientation is much faster if differential precession is so
important that it cannot be controlled by other forces
(Steiman--Cameron~\& Durisen~1988). This happens, for example, when
the external potential is such that the precession frequency is always
not very small compared to the orbital angular frequency. However,
this is not the case in the near Keplerian disks we consider. There
the precession frequency is relatively very small at all inclinations.

\subsection{Plan of the paper}

\noindent The plan of the paper is as follows. The response of the
disk to the companion perturbing potential is analyzed in
\S~\ref{sec:response}. We first give the basic equations and the
expression for the time averaged perturbing potential we later use.
We have taken the latter to be due to a companion on an inclined
orbit, but any $m=1$ perturbing potential proportional to the
coordinate $z$ along the disk rotation axis could equally well be
considered. We then describe the disk equilibrium structure, linearize
the basic equations and derive the tilt equation governing global
variations in the disk inclination taking into account self--gravity
and pressure. All effects contributing to low frequency bending wave
propagation across the disk are included to leading order.
 
\noindent The disk response and precession frequency are derived using
two different approaches, which are shown to be equivalent in the
limit of small inclination of the companion orbit.  We consider i) the
excitation of small amplitude low--frequency bending modes, considered
as perturbations of the disk away from the companion orbital
plane. Here the time averaged potential of the companion is included
in the equilibrium state. We also consider ii) the zero frequency
response of a disk forced by the time averaged potential due to a
companion in a general inclined orbit. Here the companion is not
included in the equilibrium state. We go on to discuss the effect of
viscosity, showing how this has the effect of reducing the importance
of the pressure forces in comparison to those due to self--gravity
once the viscosity parameter $\alpha$ significantly exceeds $H/r.$ We
also show that when the disk warping is small, the disk inclination
evolves on a long timescale.

\noindent In \S~\ref{sec:NGC} we apply these results to NGC~4258. We
derive the disk parameters from the observations and present the
numerical results. We find that, if the disk is self--gravitating, the
observed warp may be produced by a companion of comparable mass to
that of the disk, with an orbital separation of about one and a half
times the observed outer radius of the disk. The presence of the
companion may also cause the disk truncation. We also show that
viscous forces induce a twist in the orientation of the spin axes of
annuli at different radii. For the estimated disk parameters, only a
small trailing twist is expected to be associated with the warp if a
perturbing companion alone produces it. At the present moment, there
is no strong constraints on the magnitude of the twist. But an
accurate measure of it can provide useful information on the non
conservative forces acting in the disk.

\noindent We summarize and discuss our results in
\S~\ref{sec:discussion}.

\section{Disk response}
\label{sec:response}

\subsection{Basic equations}

The dynamics of the disk is described by the equation of motion

\begin{equation}
\frac{\partial {\bf v}}{\partial t} + \left( {\bf v} \cdot
\mbox{\boldmath $\nabla$} \right) {\bf v} = - \frac{1}{\rho}
\mbox{\boldmath $\nabla$} P - \mbox{\boldmath $\nabla$} \Psi + {\bf
f_v} ,
\label{motion}
\end{equation}

\noindent and the equation of continuity
 
\begin{equation}
\frac{\partial \rho}{\partial t} + \mbox{\boldmath $\nabla$} \cdot
\left( \rho {\bf v}
\right) = 0 ,
\label{continuity}
\end{equation}
 
\noindent where $P$ is the pressure, $\rho$ the density, ${\bf v}$ the
flow velocity and $\Psi$ the total gravitational potential. We allow
for the possible presence of a viscous force per unit mass ${\bf f_v}$
but we shall assume that it does not affect the undistorted
axisymmetric disk so that it operates on the perturbed flow only.

\noindent We write $\Psi=\Psi_{ext}+\Psi_G$, where $\Psi_{ext}$ is the
contribution to potential due to the central black hole and the
perturber and $\Psi_G$ is the contribution to the potential arising
from the disk self--gravity given by

\begin{equation}
\Psi_G = -G \int_V {\frac{\rho \left( {\bf r'} \right) d^3{\bf r'}}
{\left| {\bf r}-{\bf r'} \right|}} ,
\label{psiG}
\ee

\noindent the integral being taken over the disk volume, with ${\bf
r}$ and ${\bf r'}$ denoting position vectors and $G$ being the
gravitational constant.
 
\noindent For our calculations, we adopt a polytropic equation of
state $ P = K \rho^{1+1/n} $, $K$ and $n$ being the polytropic
constant and index respectively. Then $\Omega$ is a function of $r$
alone, and the sound speed is given by $c_s^2=dP/d\rho$.

\subsection{Perturbing potential}
\label{sec:potential}

\noindent We consider the response of the thin self--gravitating disk
to a perturbing gravitational potential that has a plane of symmetry
which is inclined to that of the disk. We suppose that the disk is
perturbed by a point mass $M_p$ orbiting as a binary companion to the
point mass $M$ located at the disk centre. In our presentation, the
disk outer radius is denoted by $R$ and the binary separation by $D.$

\noindent We consider a non rotating Cartesian coordinate system
$(x,y,z)$ centered on the central point mass $M$. The $z$ axis is
chosen to be along the rotation axis of the unperturbed disk. We shall
also use the associated cylindrical polar coordinates $(r, \varphi,
z)$. We take the orbit of the perturbing mass to be in a plane which
has an inclination angle $\delta$ with respect to the $(x,y)$ plane
and the line of nodes to coincide with the $x$ axis. We denote the
position vector of the perturbing mass by $\bf{D}$ with $D \equiv
|{\bf D}|$.
 
\noindent The potential $\Psi_{ext}$ arising from both the central and
the orbiting masses is given by
 
\begin{displaymath}
\Psi_{ext} = - \frac{GM}{\mid \bf{r} \mid} - \frac{GM_p}{\mid \bf{r} -
\bf{D} \mid} + {GM_p{\bf r} . {\bf D}\over D ^3}
\end{displaymath}

\noindent The last indirect term accounts for the acceleration of the
origin of the coordinate system. We are interested in the warping of
the disk which is excited by terms in the potential which are odd in
$z$ and which have azimuthal mode number $m=1$ when a Fourier analysis
in $\varphi$ is carried out. We consider only the secular term in the
potential, that is to say the zero--frequency term, because this is
the dominant contribution for producing the large--scale warp
structure, in which we are interested. This term is obtained by taking
a time average of the total perturbing potential, and is equivalent to
replacing the full potential by the one obtained if the perturbing
mass is distributed uniformly along the orbit. Non zero--frequency
terms would be important in determining the disk response if it
extended much beyond the inner Lindblad (2:1) resonance with the
companion. Here we suppose that truncation of the disk, through the
action of bending and other waves excited by interaction with the
companion, occurs such that $D=1.5R$ (see for example Lin~\&
Papaloizou~1993), which more or less excludes the resonances required
for effective wave excitation from the disk. Non secular bending waves
propagating into the disk interior have short wavelength and do not
give a global response. This is supported by the numerical simulations
of Larwood {\it et al.}~(1996) who found that the global precession of
truncated disks could be accounted for by the secular response.

\noindent The term in the Fourier expansion of the potential which is
of the required form is given by

\be \Psi'_{ext} = { \sin\varphi \over 4\pi^2}\int^{2\pi}_0 d ( \omega
t ) \int^{2\pi}_0 \left[ \Psi_{ext}(r,\varphi',z, \omega t)
-\Psi_{ext} (r,\varphi',-z, \omega t) \right] \sin \varphi' d\varphi'
, \ee

\noindent where $\omega$ is the companion's orbital frequency. The
parameters we shall use in the numerical calculations require a
development to eighth order in $D^{-1}$ of this integral:

\begin{equation} 
\Psi'_{ext} = - \frac{3}{4} \frac{GM_p}{A^{5/4} D^3} \; rz \; \sin 2
\delta \; \sin\varphi \; \left[ 1 + \frac{a_1}{A} \left( \frac{r}{D}
\right)^2 + \frac{a_2}{A^2} \left( \frac{r}{D} \right)^4 +
\frac{a_3}{A^3} \left( \frac{r}{D} \right)^6 + \frac{a_4}{A^4} \left(
\frac{r}{D} \right)^8 \right],
\label{ppsip}
\end{equation}

\noindent with $A=\left(1+r^2/D^2 \right)^2$ and

\begin{eqnarray}
a_1 & = & 2.1875 \left( 2 - 1.5 \sin^2 \delta \right), \nonumber \\
a_2 & = & 6.0156 \left( 3 - 4.5 \sin^2 \delta + 1.875 \sin^4 \delta
\right), \nonumber \\ a_3 & = & 18.3289 \left( 4 - 9 \sin^2 \delta +
7.5 \sin^4 \delta - 2.1875 \sin^6 \delta \right), \nonumber \\ a_4 & =
& 59.2022 \left( 5 - 15 \sin^2 \delta + 18.75 \sin^4 \delta - 10.9375
\sin^6 \delta + 2.4609 \sin^8 \delta \right) \nonumber.
\end{eqnarray}

\noindent Since we consider the thin disk limit, additional terms,
smaller by factors containing powers of $z/r,$ have been
neglected. Throughout this paper we shall use the complex potential

\be \Psi'_{ext} = i \; \frac{3}{4} \frac{GM_p}{A^{5/4} D^3} \; r f(r)
z \; \sin 2 \delta \; e^{i\varphi} ,
\label{ppsip2}
\ee

\noindent defined such that its real part is equal to the physical
potential. The function $f(r)$ is the term in brackets
in~(\ref{ppsip}).

\subsection{Equilibrium structure of the disk}

\noindent In the absence of perturbation, the disk is axisymmetric so
that in cylindrical coordinates the velocity ${\bf v}=(0,r\Omega,0)$
and the hydrostatic equilibrium equations are satisfied in the form

\begin{equation}
\frac{1}{\rho} \frac{\pl P}{\pl r}= - \frac{\pl \Psi}{\pl r}+r\Omega^2
,
\label{equr}
\end{equation}
 
\begin{equation}
\frac{1}{\rho} \frac{\pl P}{\pl z}= - \frac{\pl \Psi}{\pl z} .
\label{equz}
\end{equation}

\noindent In the absence of perturbation or orbiting companion, the
disk is under the influence of the central point mass $M$ so that
$\Psi_{ext}=-GM/\sqrt{r^2+z^2}.$ In the present context, we consider
thin disks for which the radial scale over which physical parameters
vary is very much larger than the vertical scale height. In the limit
where the Toomre parameter $Q$ is of order unity, the disk's
self--gravity, which may be evaluated as if the disk had zero
thickness, is then more important by a factor of order $r/H$ than
pressure in~(\ref{equr}). The angular velocity is given by
equation~(\ref{equr}) as

\begin{equation}
\Omega^2 = \Omega_K^2 + \frac{1}{r \rho} \frac{\pl P}{\pl r} +
\frac{1}{r} \frac{\pl \Psi_G}{\pl r} ,
\label{omega}
\end{equation}

\noindent where $\Omega_K=\left({GM/r^3}\right)^{1/2}$ is the
Keplerian angular velocity. Typically the contribution of
self--gravity gives $\Omega-\Omega_K= O(H\Omega_K r^{-1}).$ As a
check, we evaluated the contribution of the pressure term for the
models we consider here and verified that it was relatively small.

\noindent We could also add the time average potential due to a
companion with small mass, which orbits in the disk plane, into
$\Psi_{ext}$ and incorporate it in the equilibrium. This would give a
contribution of order $\omega_z,$ being the orbital precession
frequency of a free disk particle, to $\Omega-\Omega_K.$ However, we
shall limit ourselves to companion masses small enough so that
$|\omega_z| \ll (H\Omega r^{-1})$ and its effect may in general be
considered to be small in comparison to that due to self--gravity.
But, in a formal sense, we shall consider that typically $|\omega_z|
\gg (H^2\Omega r^{-2}),$ so that it may be retained while frequencies
of order $H^2\Omega r^{-2}$ may be neglected.

\noindent The warping and precession of a disk at small inclination
may be considered as a small amplitude perturbation about such a state
incorporating a companion (e.g. Toomre~1983; Sparke~1984) and see
below.

\noindent In principle, the surface density distribution, $\Sigma=
\int^{\infty}_{-\infty} \rho dz,$ is determined by the viscous
evolution of the disk. In the absence of a deterministic prescription
for the effective viscosity, for illustrative purposes we arbitrarily
choose $\Sigma(r)= \Sigma_0 \left( R/r -1 \right),$ $\Sigma_0$ being a
constant. Alternative functional forms describing the way $\Sigma$
tapers to zero have also been considered below.

\subsection{Linear perturbations}

\label{sec:perturb}

\noindent In the limit where either any perturbing/forcing potential
is small compared to that due to the central mass and the amplitude of
any free bending modes is small, the disk response can be described in
a linear analysis. We here follow the standard procedure of vertical
averaging (see, for example, Hunter~\& Toomre~1969, Sparke~1984,
Papaloizou~\& Lin~1995). This should be valid when the radial
wavelength of the disturbance is significantly longer than the disk
thickness as is the case for the perturbations considered here. For
linear warps, the dependence on $\varphi$ and $t$ may be taken to be
through a factor exp$[i(m \varphi + \sigma t)].$ Henceforth this
factor will be taken as read. For the time being we shall leave it as
a multiplicative factor for all perturbations but will remove it at a
later stage. The mode frequency, $\sigma,$ can be taken to be zero for
a secular response viewed in an appropriate rotating reference frame
and we shall consider this case below. Alternatively, the disturbance
may appear with a small non zero frequency or pattern speed when
viewed in an inertial frame. In the latter case, the disk will appear
to precess if the perturbation has a global form. We limit ourselves
to considering frequencies significantly less than the inverse density
wave crossing time of the disk (characteristically $O(H\Omega
r^{-1})$) which is a necessary condition if the disk is to exhibit
global precession (Hofner~\& Sparke~1994, Papaloizou~\& Terquem~1995).
Then we show below that, for modest warping of the disk, these two
descriptions are equivalent. Finally, we remark that the azimuthal
mode number $m=1$ throughout.

\noindent We denote the Lagrangian displacement as
${\mbox{\boldmath{$\xi$}}}\equiv (\xi_r,\xi_{\varphi},\xi_z).$ Here we
shall not assume at the outset that the horizontal components of the
displacement are zero, as it was shown in Papaloizou~\& Lin~(1995)
that these can produce significant effects in a self--gravitating but
near Keplerian disk of the type considered here.

\noindent The inclination (vertical displacement/radius) associated
with the tilt of the disk is $\xi_z/r.$ This is in general assumed to
be a slowly varying function of both $r$ and $z$ such that its
variation with $z$ may be neglected. The applicability of this
vertical averaging approximation is supported by the numerical
calculations of Papaloizou~\& Lin~(1995) which allowed for the
possibility of variation of the vertical displacement associated with
modes of the type we consider with $z.$

\noindent We first assume that we are working in an inertial frame in
which the unperturbed disk appears steady. However, it may
subsequently be convenient to adopt a uniformly rotating frame so as
to remove the rigid body precession associated with the motion of the
disk, in which case the effects due to the coriolis force need to be
added.

\noindent We begin by writing down the perturbed form of the vertical
component of the equation of motion applicable to a gaseous disk with
a barotropic equation of state (see Hunter~\& Toomre~1969)

\begin{equation}
{D^2 \xi_z \over Dt^2}= -{\partial \Psi_G' \over \partial z} -
{\partial \left( P'/ \rho \right) \over \partial z} - {\partial
\Psi_{ext}' \over \partial z},
\label{PT1} 
\ee 

\noindent where perturbations to quantities are denoted with a prime
and the convective time derivative $D/Dt$ is here equivalent to
multiplication by $i(\sigma + \Omega).$

\noindent The potential perturbation, $\Psi'_{ext}$, is taken to be
the secular contribution due to the companion in an inclined circular
orbit (see \S~\ref{sec:potential}) where this has not been included in
the equilibrium. Then we calculate the disk response taking $\sigma
=0,$ but will need to transform to a rotating frame (see below). If
the companion orbits in the initial symmetry plane of the disk and its
effect included in the equilibrium, then we look for free bending
modes with small frequency $O(\omega_z),$ the amplitude of which will
give the inclination of the precessing disk. These two approaches are
equivalent for small inclinations and modest warping.

\noindent For a polytropic equation of state we also have $P'=\rho'
c_s^2.$ Using the above, equation~(\ref{PT1}) may be written 

\begin{equation}
(\sigma + \Omega)^2 \xi_z= {\partial \Psi'_G \over \partial z} +
{\partial \left( \rho' c_s^2/ \rho \right) \over \partial z}
+{{\partial \Psi'_{ext}}\over {\partial z}}.
\label{PT2} 
\ee

\noindent The perturbation to the gravitational potential of the disk
is given by the Poisson integral

\be \Psi_G' = -G \int_V {\frac{\rho' \left( {\bf r'} \right) d^3{\bf
r'}} {\left| {\bf r}-{\bf r'} \right|}} ,
\label{PT3}
\ee

\noindent while the perturbed continuity equation gives 

\be \rho'= - \mbox{\boldmath $\nabla$} \cdot \left( \rho
\mbox{\boldmath{$\xi$}} \right).
\label{contpert} \ee

\noindent We then find after an integration by parts, assuming that
the disk density vanishes at its boundaries, that

\be \Psi'_G = -G \int_V {\frac{\rho \left( {\bf r'} \right)
\mbox{\boldmath{$\xi$}} \left( {\bf r'} \right) \cdot \left( {\bf
r}-{\bf r'} \right) d^3{\bf r'}} {\left| {\bf r}-{\bf r'} \right|^3}}.
\label{PT4}
\ee

\noindent In what follows, we find it convenient to consider the
contributions to $\rho'$ and $\Psi'_G$ arising from the horizontal and
vertical components of $\mbox{\boldmath{$\xi$}}$ separately. The
contributions arising from the vertical component we denote with a
subscript $v$ and call them vertical contributions. The contributions
arising from the horizontal components we denote with a subscript $h$
and call them horizontal contributions. Thus

\begin{eqnarray} 
\rho'_v & = & -{\partial (\rho \xi_z) \over \partial z} ,
\label{PT7} \\
\rho'_h & = & - \mbox{\boldmath $\nabla$} \cdot (\rho
{\mbox{\boldmath{$\xi$}}}_h).
\label{PT7a} \end{eqnarray}

\noindent $\Psi'_{Gv}$ and $\Psi'_{Gh}$ denote the potential arising
from $\rho'_v$, and $\rho'_h$ respectively. Thus

\begin{eqnarray} 
\Psi'_{Gv} & = & -G \int_V {\frac{\rho \left( {\bf r'} \right) \xi_z
\left( {\bf r'} \right) \left( z-z' \right) d^3{\bf r'}} {\left| {\bf
r}-{\bf r'} \right|^3}} ,
\label{PT4a} \\
\Psi'_{Gh} & = & -G \int_V {\frac{\rho \left( {\bf r'} \right)
{\mbox{\boldmath{$\xi$}}}_h\left( {\bf r'} \right) \cdot \left( {\bf
r}-{\bf r'} \right) d^3{\bf r'}} {\left| {\bf r}-{\bf r'} \right|^3}}.
\label{PT4b}
\end{eqnarray}

\noindent The vertical and horizontal contributions give rise to
additive contributions, $F'_v $ and $F'_h $ respectively, to the
vertical component of the equation of motion~(\ref{PT2}). We consider
these in turn below.

\subsubsection{ The tilt equation}

\noindent Having evaluated the vertical and horizontal contributions
we may construct a tilt equation governing the inclination $g$ derived
from vertical integration of equation~(\ref{PT2}) in the form

\be (\sigma + \Omega)^2 \; \Sigma \; \xi_z = \int^{\infty}_{-\infty}
{\rho \left( F'_h + F'_v \right) dz + \int^{\infty}_{-\infty} \rho \;
{\partial \Psi'_{ext} \over \partial z} \;dz}.
\label{PT15} \ee

\subsubsection{The vertical contribution}

\noindent From equation~(\ref{PT2}), this is seen to give rise to the
vertical deceleration 

\be F'_v = {\partial \Psi'_{Gv} \over \partial z} + {\partial \left(
\rho'_v c_s^2 / \rho \right) \over \partial z}. \ee

\noindent To deal with this, we differentiate the vertical
contribution to the Poisson integral~(\ref{PT4a}) with respect to $z,$
differentiate the Poisson integral~(\ref{psiG}) twice with respect to
$z$ and combine with $\xi_z$ to obtain in the limit of an arbitrary
thin disk in its midplane 

\be {\partial \Psi'_{Gv} \over \partial z} + \xi_z {\partial^2
\Psi_{G} \over \partial z^2} \rightarrow -G \int_A {\frac{ \Sigma
\left( {\bf r'} \right) \left[ \xi_z \left( {\bf r'} \right) - \xi_z
\left( {\bf r} \right) \right] d^2{\bf r'}} {\left| {\bf r}-{\bf r'}
\right|^3} \; },
\label{PT5}
\ee

\noindent where the two dimensional integral is over the surface of
the disk.

\noindent Proceeding further, we remark here that we shall eventually
require 

\be \int^{\infty}_{-\infty} \rho F'_v dz = -\int^{\infty}_{-\infty}
{\rho \left( \xi_z {\partial^2 \Psi_{G} \over \partial z^2} + G \int_A
{{\Sigma \left( {\bf r'} \right) \left[ \xi_z \left( {\bf r'} \right
)- \xi_z \left( {\bf r} \right) \right] d^2{\bf r'}} \over {\left|
{\bf r}-{\bf r'} \right|^3}} \right) dz} - \int^{\infty}_{-\infty}
{\rho \; {\partial \over \partial z} \left( { c_s^2 \over \rho}
{\partial \left( \rho \xi_z \right) \over \partial z} \right) dz} .
\label{PT6} \ee

\noindent Here we have made use of~(\ref{PT5})
and~(\ref{PT7}). Equation~(\ref{PT6}) may be further simplified by
using the assumption that $\xi_z$ is independent of $z,$ together with
use of the fact that the equilibrium quantities satisfy hydrostatic
equilibrium~(\ref{equz}) in the vertical direction, with the result
that

\be \int^{\infty}_{-\infty} \rho F'_v dz = \Sigma \; \xi_z \;
{\partial^2 \Psi_{ext} \over \partial z^2} - G \int_A {{\Sigma \left(
{\bf r} \right) \Sigma \left( {\bf r'} \right) \left[ \xi_z \left(
{\bf r'} \right) - \xi_z \left( {\bf r} \right) \right] d^2{\bf r'}}
\over {\left| {\bf r}-{\bf r'} \right|^3}} \; .
\label{PT8} \ee

\noindent Here and below, we make the assumption that the second
derivatives of the external potential are slowly varying through the
disk thickness so they can be evaluated in the midplane. This
completes the reduction of the vertical contribution to the force per
unit area which now depends only on the vertical displacement.

\subsubsection{The horizontal contribution}

\noindent From equation~(\ref{PT2}), this is seen to give rise to the
horizontal deceleration 

\be F'_h = {\partial \Psi'_{Gh} \over \partial z} + {\partial \left(
\rho'_h c_s^2/\rho \right) \over \partial z} . \ee

\noindent Noting again that we need to perform a vertical average, we
multiply by $\rho$ and integrate over the disk thickness to obtain
after an integration by parts of the second term and use of vertical
hydrostatic equilibrium 

\be \int^{\infty}_{-\infty} \rho F'_h dz = \int^{\infty}_{-\infty}
\left( \rho \; {\partial \Psi'_{Gh} \over \partial z} + \rho'_h \;
{\partial \Psi_{G} \over \partial z} \right) dz +
\int^{\infty}_{-\infty} \rho'_h \; {\partial \Psi_{ext} \over \partial
z}dz.
\label{PT9} \ee

\noindent We pause for a moment to consider the magnitudes of the
terms in equation~(\ref{PT9}). Referring back firstly to
equation~(\ref{PT8}), we comment that the first term on the right hand
side there is of order $\Omega_K^2 \Sigma \xi_z$ while the second,
being produced by the self--gravity of the disk, is characteristically
smaller by a factor $H/r$ for Toomre parameter $Q$ of order unity and
global perturbations as considered here. Allowing for the possibility
that the horizontal displacement could become as large in magnitude as
the vertical one, the first two terms in equation~(\ref{PT9}), being
produced by the disk self--gravity, are each also of potential
magnitude $\Omega_K^2 \Sigma H \xi_z /r$. However, we show in the
Appendix that, for global perturbations, the combination is in fact
smaller by an additional factor of order $H/r,$ and so can be
neglected in our ordering scheme. Thus only the final term remains in
equation~(\ref{PT9}). This is best dealt with by considering the
perturbed equations of motion in Eulerian form. These can be written
(remembering $m=1$)

\begin{eqnarray}
i \left( \sigma+\Omega \right) v'_r - 2 \Omega v'_{\varphi} & = & -
{\partial W \over \partial r}, \label{PT10} \\ i \left( \sigma +
\Omega \right) v'_{\varphi} + {\kappa^2 \over 2 \Omega} v'_r & = &
-{iW\over r}, \label{PT11} \\ i \left( \sigma+\Omega \right) v'_z & =
& - {\partial W \over \partial z}. \label{PT12} \end{eqnarray}

\noindent Here the velocity perturbation is ${\bf v}'= (v'_r,
v'_{\varphi}, v'_z),$ $W= P'/\rho +\Psi',$ and $\kappa^2=2 \Omega
r^{-1} d \left( r^2 \Omega \right) /dr$ is the square of the epicyclic
frequency.

\noindent Given that we seek an inclination $ \xi_z/r \equiv v'_z/
\left[ i r \left( \sigma+\Omega \right) \right]$ that is independent
of $z,$ we may write $W= \left( \sigma+\Omega \right)^2 z \xi_z.$
Equations~(\ref{PT10}) and~(\ref{PT11}) may then be used to solve for
the horizontal velocity perturbations in terms of $\xi_z.$ These can
then be used to find $\rho'_h$ from the perturbed continuity
equation~(\ref{contpert}). An analagous procedure has been carried out
in Papaloizou~\& Lin~(1995) and Papaloizou~\& Terquem~(1995). As we
are here interested in the low frequency limit, it is consistent with
our ordering scheme to perform this procedure adopting $\sigma=0.$ We
thus obtain

\begin{eqnarray} 
v'_r & = & {-i \Omega z \over r^2 \left( \kappa^2-\Omega^2
\right)}{\partial \left( \xi_z r^2 \Omega^2\right) \over \partial r},
\label{vrp} \\
v'_{\varphi} & = & {\kappa^2 z \over 2 r^2 \Omega \left( \kappa^2 -
\Omega^2 \right)}{\partial \left( \xi_z r^2 \Omega^2 \right) \over
\partial r} -{z \xi_z \Omega \over r}. \label{vphip} \end{eqnarray}

\noindent Finally we obtain from~(\ref{contpert})

\be \rho'_h=-{1 \over i \Omega} \mbox{\boldmath $\nabla$} \cdot \left(
\rho {\bf v}'_h \right) = {z \over \Omega r^{1/2}} {\partial \over
\partial r} \left( { \rho \Omega \over r^{3/2} \left(
\kappa^2-\Omega^2 \right)} {\partial \left( \xi_z r^2 \Omega^2 \right)
\over \partial r}\right). \label {PT13} \ee

\noindent In forming the above, consistently with the neglect of
frequencies of order $H^2\Omega_K r^{-2},$ we have neglected the
contribution of the last term in (\ref{vphip}) to (\ref{PT13}).

\noindent Using equation~(\ref{PT9}), we find the horizontal
contribution to the vertically integrated vertical component of the
equation of motion

\be \int^{\infty}_{-\infty} \rho F'_h dz = {\partial^2 \Psi_{ext}
\over \partial z^2} {1 \over \Omega r^{1/2}} {\partial \over \partial
r} \left( { \mu \Omega \over r^{3/2} \left( \kappa^2-\Omega^2 \right)}
 {\partial \left( \xi_z r^2\Omega^2 \right) \over \partial
r}\right), \label{PT14} \ee 

\noindent where we have approximated $ \partial \Psi_{ext}/ \partial z
= (\partial^2 \Psi_{ext}/ \partial z^2)z,$ the second derivative being
evaluated on the midplane, and $\mu = \int_{-\infty}^{\infty} \rho z^2
dz.$

\subsubsection{Reduction of the  tilt equation}

\noindent Having evaluated the vertical and horizontal contributions,
we construct the tilt equation~(\ref{PT15}) which now appears as an
equation for $\xi_z$ alone.

\noindent Using equations~(\ref{PT8}) and~(\ref{PT14}), we obtain

\begin{displaymath}
\left[ \left( \sigma + \Omega \right)^2 - {\partial^2 \Psi_{ext} \over
\partial z^2} \right] \Sigma \; \xi_z + G \int_A {{\Sigma \left( {\bf r}
\right) \Sigma \left( {\bf r'} \right) \left[ \xi_z \left( {\bf r'}
\right)- \xi_z \left( {\bf r} \right) \right] d^2 {\bf r'}}\over
{\left| {\bf r}-{\bf r'} \right|^3}} 
\end{displaymath}

\be = r \; \Omega_K^2 \; {\partial \over \partial r} \left({ \mu
\Omega_K^2 \over \kappa^2-\Omega^2}  {\partial \left( \xi_z/r
\right) \over \partial r}\right) + \Sigma \; {\partial \Psi'_{ext} \over
\partial z} .
\label{PT16} \ee

\noindent Here, consistently with the neglect of frequencies of order
$H^2\Omega_K r^{-2},$ we have made the replacements $ \partial^2
\Psi_{ext} / \partial z^2 =\Omega^2=\Omega_K^2$ in~(\ref{PT14})
everywhere apart from in the denominator $ \kappa^2-\Omega^2$.  Also
in the last term, the derivative is evaluated on the midplane.

\noindent Rather than retaining $\xi_z$ as the quantity to be
determined, we adopt $g=i \left( \xi_z/r \right) \exp \left( -i
\varphi \right).$ At zero frequency, its modulus is the ratio of
vertical to azimuthal velocity and also the disk inclination.
Remembering that the azimuthal dependence of $\xi_z$ is through a
factor $\exp \left( i \varphi \right),$ the replacement with $g$ which
is accordingly independent of $\varphi$ will remove this factor from
equation~(\ref{PT16}). Carrying out this reduction we obtain

\be \left[ \left( \sigma + \Omega \right)^2 - \frac{1}{r} {\partial
\Psi_G \over \partial r} - {\partial^2 \Psi_{ext} \over \partial z^2}
\right] {\Sigma g \over \Omega_K^2} = {\cal L}\left( g \right) + {i
\Sigma \over r \Omega_K^2} {\partial \Psi'_{ext} \over \partial z} ,
\label{PT17} \ee

\noindent where the operator ${\cal L}$ is defined through 

\be {\cal L} \left( g \right) = \frac{G}{r^3 \Omega_K^2} \int_{R_i}^R
\hat{K} \left( r,r' \right) \Sigma(r) \Sigma(r') \left[ g(r) - g(r')
\right] r'^2 r^2 dr' + {\partial \over \partial r} \left( { \mu
\Omega_K^2 \over \kappa^2-\Omega^2 } {\partial g \over \partial
r}\right) . \label{PT18} \ee

\noindent Here the kernel $\hat{K}$ is given by 

\be \hat{K}(r,r')= \int_0^{2 \pi} \frac{\cos \Phi d\Phi} {\left[ r^2 +
r'^2 - 2rr' \cos \Phi \right] ^{3/2}} , \ee

\noindent and $R_i$ is the innermost disk radius.

\noindent We also note that after using (\ref{equr}), for the low
frequencies we consider, to the accuracy we are working

\begin{displaymath}
\Omega^2 - \frac{1}{r} {\partial \Psi_G\over \partial r}- {\partial^2
\Psi_{ext}\over \partial z^2} = \frac{1}{r} {\partial \Psi_{ext}\over
\partial r} -{\partial^2 \Psi_{ext}\over \partial z^2} = 2 \Omega_K
\omega_z,
\end{displaymath}

\noindent where again $\omega_z(r)$ is the orbital precession
frequency due to the companion if it orbits in the symmetry plane and
is included in the equilibrium external potential so that
$\Psi'_{ext}=0.$ If instead the effect of a companion in an inclined
orbit is included only through $\Psi'_{ext},$ we have $\omega_z=0.$

\noindent Neglecting $\sigma^2,$ the inviscid tilt equation is then
written in the convenient form:

\be {2 \left( \sigma + \omega_z \right) \Sigma g \over \Omega_K} =
{\cal L}(g) + {i \Sigma \over r \Omega_K^2} {\partial \Psi'_{ext}
\over \partial z} . \label{PT19} \ee

\noindent We comment that the above equation describes the global
response of a thin near Keplerian self--gravitating disk with Toomre
parameter $Q$ of order unity. The largest frequencies neglected are of
order $H^2\Omega_k r^{-2}.$ It can also be used to describe low
frequency bending modes when the forcing term is set to zero. There
are two identifiable contributions to the operator ${\cal L}$ defined
by~(\ref{PT18}). The first comes from self--gravity and leads to the
description of warps given by Hunter~\& Toomre~(1969), Sparke~(1984),
Sparke~\& Casertano~(1988), Kuijken~(1991) and others. The second term
in ${\cal L}$ can be identified as arising from pressure and it can
lead to comparable effects to those due to self--gravity in a near
Keplerian disk, in contrast to a galactic disk, because of the near
Lindblad resonance that arises in the former case because
$\kappa-\Omega=O(H\Omega r^{-1}).$ The form of this term is the zero
frequency global counterpart of that occurring in the low frequency
WKB dispersion relations given by Papaloizou~\& Lin~(1995) and
Masset~\& Tagger~(1996).

\noindent Another useful property noted by Sparke~\& Casertano~(1988)
in the pure self--gravity problem is that, for a finite isolated disk,
${\cal L}$ is self--adjoint, that is for any pair $(g_1, g_2)$ that
satisfy appropriate regularity conditions, including zero derivative
at a Lindblad resonance if the pressure term is included:

\be \int^R_{R_{i}} g_2^{*} {\cal L} \left( g_1 \right) dr = \left(
\int^R_{R_{i}} g_1^* {\cal L} \left( g_2 \right) dr \right)^*. \ee

\noindent Furthermore, when $\omega_z=0$ and there is no forcing,
equation~(\ref{PT19}) has the solution $g = {\rm constant}$ which
coresponds to a rigid tilt in a pure Keplerian potential. Modification
of this rigid tilt mode by a companion results in a low frequency
modified tilt mode that describes global warping and precession of the
disk.

\subsubsection{Disk precession}

\label{sec:precession}

\noindent {\it i) Low frequency bending modes}

\noindent We first consider the case of a disk precessing at low
inclination about the angular momentum vector associated with the
companion orbit. This can be viewed as the excitation of a small
amplitude bending mode with the time averaged potential of the
companion being included in the equilibrium state. Implicit in this
treatment is the notion that the companion orbit is fixed so its
angular momentum content must be presumed to be significantly larger
than that of the disk. Such modes are governed by the eigenvalue
problem

\be {2 \left( \sigma + \omega_z \right) \Sigma g \over \Omega_K} =
{\cal L}(g) . \label{PT20} \ee

\noindent When $\omega_z =0,$ there is a solution corresponding to the
rigid tilt mode with $g= g_0 ={\rm constant}=\delta,$ where for
convenience we take $\delta$ to be real and therefore equal to the
inclination and $\sigma=0.$ For small $\omega_z,$ there is a solution
with $g$ close to $g_0$ and $\sigma$ non zero but small in
magnitude. From first order perturbation theory, remembering ${\cal
L}(g_0)=0,$ we obtain the forced response problem for $g$:

\be {\cal L}(g) = {2 \left( \sigma + \omega_z \right) \Sigma \delta
\over \Omega_K} . \label{PT21} \ee

\noindent Multiplying by $g_0$, integrating over the disk and using
the self--adjoint property of ${\cal L}$, we find the integrability
condition which determines $\sigma$:

\be \sigma\int^{R}_{R_{i}}{\Sigma g_0  \delta \over \Omega_K}dr=
-\int^{R}_{R_{i}}{\Sigma \omega_z g_0 \delta  \over \Omega_K}dr .\ee
Remembering that $g_0$ is constant and the form of Kepler's law, this
may be written

\be \sigma = - \int^{R}_{R_{i}} 2\pi r \Sigma(r) \omega_z(r) j(r)
\delta dr \left/ \int^{R}_{R_{i}} 2\pi r \Sigma(r) j(r) \delta dr
\right. , \ee 

\noindent where $j(r)= r^2\Omega_K$ denotes the specific angular
momentum of the orbiting disk material.

\noindent This relation has a simple interpretation, as the gyroscope
equation, that the magnitude of the disk precession frequency
$-\sigma$ should equal the component of the applied torque
perpendicular to the companion angular momentum axis and the disk
angular momentum axis divided by the magnitude of the component of the
disk angular momentum vector perpendicular to the companion angular
momentum axis (Kuijken~1991). This relation is applicable to secular
precession of free particle orbits and thus determines $\omega_z,$ so
$\omega_zj(r)\delta= - T(r),$ where $T(r)$ is the external torque
component per unit mass acting on a disk circular orbit at radius $r.$

\noindent Once the precession frequency has been found, the forced
response problem can be solved to find the degree of warping. For the
idea of global precession to be valid this should be small. This
condition is roughly equivalent to the requirement that the disk wave
crossing time be short compared to the precession period.

\noindent Another usefull relation that can be obtained
from~(\ref{PT20}) is found by multiplying by $g^*$ and integrating
over the disk, giving

\be 2\sigma \int {\Sigma g^* g \over \Omega_K} dr +2 \int {\omega_z
\Sigma g^* g \over \Omega_K}dr = \int g^*{\cal L}(g)dr . \label{PT20A}
\ee

\noindent The eigenfrequency $\sigma,$ if real, corresponds to a
simple precession. However, if it were complex, the imaginary part
would give the rate of decay of the inclination towards the symmetry
plane. From~(\ref{PT20A}) we obtain for this decay rate

\be Im (\sigma) = Im \left( \int g^* {\cal L}(g) dr \right) \left/
\left( 2 \int \frac{\Sigma g^* g}{\Omega_K} dr \right) \right. .
\label{PT20B} \ee

\noindent Because of the self--adjoint property of the problem
considered up to now, the decay rate is zero. However, if
contributions from viscous or other non conservative forces are added,
a non zero decay rate may result (see below).

\noindent {\it ii) External forcing at finite inclination}

\noindent In this case, we consider the zero frequency response of a
disk forced by the secular potential due to a companion in an inclined
orbit given by (\ref{ppsip2}) so that $\omega_z=\sigma=0,$ and the
governing equation is

\be {\cal L}(g) = -{i \Sigma \over r \Omega_K^2} {\partial \Psi'_{ext}
\over \partial z}.  \label{PT22} \ee

\noindent But because the unforced problem has $g={\rm constant}$ as a
solution, the forced problem will not in general have a
solution. Physically, this is because there is an unbalanced torque
due to the companion which produces disk precession. To deal with
this, we suppose the disk appears steady in a frame precessing about
the total angular momentum axis with frequency $\omega_p.$ At first,
let us suppose this coincides with the companion orbital angular
momentum vector as in the preceeding section. Then we must add the
vertical component of the coriolis force as a perturbing force along
with that due to the external potential. This amounts to an additional
perturbing force per unit mass being the real part of $-2 i\omega_p
\sin(\delta) r \Omega_K \exp \left( i\varphi \right).$ Incorporating
this into equation~(\ref{PT21}) (i.e. adding it to $ -\partial
\Psi'_{ext}/ \partial z$) results in this now becoming

\be {\cal L}(g) = { \Sigma \over r \Omega_K^2} \left( 2 \omega_p r
\Omega_K \sin\delta -i {\partial \Psi'_{ext} \over \partial z} \right)
. \label{PT23} \ee

\noindent The above problem is very similar to, and in fact equivalent
to, (\ref{PT21}) for small $\delta.$ Just as in that case, we have a
forced response problem to solve for $g$ and we find the integrability
condition that determines $\omega_p$ to be

\be
\omega_p = \int_{R_i}^R \frac{i \Sigma}{2r \Omega_K^2}
\left( \frac{\pl \Psi'_{ext}}{\pl z} \right) dr
\left / \int_{R_i}^R \frac{ \Sigma \sin \delta}{\Omega_K } dr .
\right.
\label{prec}
\ee

\noindent This equation can also be interpreted as a gyroscope
equation giving the precesion frequency as the ratio of appropriate
disk torque and angular momentum components. Note here that the
perturbing external potential is complex (see \ref{ppsip2}) ensuring
that the calculated precession frequency is real and that the factor
of two in (\ref{prec}) accounts for an azimuthal average.

After finding $\omega_p$ from~(\ref{prec}), equation~(\ref{PT23}) may
be integrated to give $g,$ to within the addition of an arbitrary
constant inclination, which may be eliminated by choosing the
coordinate system so that $g=0$ at the disk inner boundary. Then if
$g$ is small elsewhere, the disk approximately precesses like a rigid
body.

\noindent We have mentioned in \S~\ref{sec:settle2} that the disk is
not expected to have settled to equilibrium with zero inclination to
the orbital plane. The choice of zero inclination perturbation at the
inner boundary is then justified.

\noindent From~(\ref{ppsip}), we see that $\omega_p \propto
\cos\delta$ when $D \gg R$ such that the tidal potential becomes
quadrupolar. The same dependance is found for a disk subject to the
potential of a misaligned flattened halo, $\delta$ being in that case
the angle between the disk plane and the symmetry plane of the
potential (Kuijken~1991).

\noindent Up to now we have assumed that the disk angular momentum
content can be neglected in comparison to that of the orbiting
companion. Then the disk precesses about the companion angular
momentum axis.  For finite disk angular momentum content, the disk and
companion both precess about the conserved total angular momentum
vector.  This can be taken account of by noting that the precession
frequency $\omega_p$ as calculated above should be replaced by
$\omega_p J_{orb}/J$, where $J_{orb}$ is the companion's orbital
angular momentum and $J$ is the total angular momentum.

\subsubsection{The effect of viscosity}
\label{sec:viscosity}

\noindent In principle viscosity can act on the disk so as to change
its inclination with respect to the orbital plane (see
\S~\ref{sec:settle2}), and also by damping the horizontal velocities
that can be amplified by a close Lindblad resonance (see
equations~\ref{vrp} and~\ref{vphip}).

\noindent We have already observed in \S~\ref{sec:settle2} that, as
long as it is globally warped, the timescale on which the disk
inclination is expected to change is the global viscous timescale,
namely $\Omega^{-1} (R/H)^2/\alpha$, where $\alpha$ is the standard
Shakura~\& Sunyaev~(1973) parameter. Since $\alpha \le 1$, evolution
on this timescale is long and has been neglected in comparison to the
other processes considered in the above analysis, where frequencies on
the order of $H^2 \Omega r^{-2}$ have been neglected. Thus, in the
ordering scheme we have adopted, changes in the disk inclination with
respect to the orbital plane should be negligible.

\noindent In a near Keplerian disk, viscosity also provides a damping
effect on the near resonantly produced horizontal velocity
perturbations. We follow the discussion given by Papaloizou~\&
Lin~(1994, 1995). The near resonant denominator acting in
equations~(\ref{vrp}) and~(\ref{vphip}) is $1- \kappa^2 /\Omega^2 \sim
O(H/r)$ when $Q\sim 1.$ The effect of an appropriately defined
$\alpha$ viscosity, at the order we are working, is to replace this
denominator by $(1-i\alpha)^2- \kappa^2 /\Omega^2.$ Demianski~\&
Ivanov~(1997) have recently given the relativistic
generalization. Thus viscosity can only be neglected when $\alpha \ll
H/r$, which is not the case in some models of NGC~4258 (see below).

\noindent As indicated above, we take viscosity into account by
changing the denominator $\kappa^2-\Omega^2$ of the last term of the
operator ${\cal L}$ (equation~\ref{PT18}) to
$\kappa^2-\Omega^2(1-i\alpha)^2$. We note that when $\alpha \gg H/r$,
the pressure term in ${\cal L}$ is small compared to that due to
self--gravity, so that the warp is mainly controlled by
self--gravity. This is illustrated below in the case of models for
NGC~4258.

\noindent We comment that incorporation of viscosity as described
above does not affect the discussion leading to the determination of
the precession frequency through equation (\ref{prec}). This is
consistent with the idea that as long as the disk is only mildly
warped, the internal viscous forces do not enable a net torque to be
exerted on the disk, and so do not modify the precession rate or
change the disk inclination except possibly on a long timescale.

\noindent In the case of small inclination we may use (\ref{PT20B}) to
find the rate of decay of the inclination after incorporating the
effect of viscosity as outlined above. For this we then obtain
 
\be Im (\sigma) = \int { 2 \alpha \mu \Omega^2 \Omega_K^2 \over \left[
\kappa^2 - \Omega^2 \left( 1-\alpha^2 \right) \right]^2 + 4 \alpha^2
\Omega^4 } \left| {\partial g \over \partial r} \right|^2 dr \left/
\left( 2 \int \frac{\Sigma g^* g}{\Omega_K} dr \right) \right. .
\label{PT20M} \ee

\noindent At a given location in the disk, the decay rate is a maximum
for $\alpha\sim H/r.$ For $\alpha > H/r,$ the decay rate is found to
be of order $\Omega(H/r)^2\alpha \left[ \delta g/(\alpha g)
\right]^2,$ where $\delta g$ represents the total change in
inclination or the total degree of warping in the system. Thus the
decay rate is small if the total warp is small.

\section{Application to NGC~4258}
\label{sec:NGC}

A warped disk model was originally proposed for NGC~4258 to account
for the origin of the megamasers (Neufield~\& Maloney~1995). In this
scenario, the warp provides a favorable condition for parts of the
disk to be illuminated by the X--ray source within the nucleus.  In
the regions between $0.13-0.26$~$pc$, the surface layers of the disk
are heated to temperatures in the range $300$~$K < T <8,000$~$K$ at
which water maser production may occur (Collison~\&
Watson~1995). Neufield~\& Maloney~(1995) supposed that conditions are
less favourable for reprocessing in those regions of the disk interior
to $0.13$~$pc$ and exterior to $0.26$~$pc$.

With respect to the systemic velocity of NGC~4258, the flux of the
red--shifted masers is an order of magnitude larger than that of the
blue--shifted masers. Herrnstein {\it et al.}~(1996) suggested that
the heated gas above the concave side of a warped disk leads to
greater extinction than that above the convex side.  With the
appropriate orientation, thermal absorption is dominant for the
blue--shifted masers.

\subsection{Disk Parameters}
\label{sec:parameters}

For the total luminosity of $L=4\times 10^{41}$~$erg/s$ attributed to
this object, an efficiency $\epsilon$ (here taken to be 0.1) of rest
mass conversion implies a mass accretion rate through the system of

\be
\dot M = {L\over \epsilon c^2}= 6\times 10^{-5} \; M_{\odot} /yr.
\label{MDOT}
\ee

Note that the need for reprocessing is evident as according to a
steady state optically thick disk model (see Pringle~1981) the
effective temperature would be given by

\begin{displaymath}
T_{eff}=11(r_{18})^{-0.75} \; K,
\end{displaymath}

\noindent where $r_{18}$ is the radius $r$ in units of $10^{18}$~$cm$.
For comparison, the effective temperature, assuming that the central
luminosity is radiated over a radial scale $r_{18}$, would be

\begin{displaymath}
T_{eff}=154(r_{18})^{-0.5} \; K.
\end{displaymath}

\noindent Thus, in the most region of the disk, the irradiation is the
dominant source of heating and $T_{eff}$ is in the range appropriate
for water maser emission.

\noindent The disk semi--thickness $H$ is given by

\begin {equation}
{H\over r}\sim {c_s\over r\Omega} \sim 1.5\times
10^{-3}\left({T_{200} r_{18}\over
M_{7.6}}\right)^{1/2}.
\label{THICK}
\end{equation} 

\noindent Here $M_{7.6}= M/(4\times 10^{7}$~$M_{\odot}),$ $M$ is the
central black hole's mass, $c_s$ is the sound speed,
$T_{200}=T_c/200$~$K$, $T_c$ is the midplane temperature, and $\Omega$
is the orbital rotational frequency in the disk. From the observed
upper limit $H/R < 2.5 \times 10^{-3}$, we find $T_c \sim T_{eff}$ due
to irradiation and $T_{200} r_{18}/ M_{7.6} < 2.8.$

\noindent Applying the conventional $\alpha$ prescription for an
effective turbulent viscosity (Shakura~\& Sunyaev~1973), the mass
transfer rate in a steady--state disk is 

\be{\dot M}= 3\pi\alpha H^2\Omega\Sigma= 0.59\alpha
\frac{M_D}{M} \left({T^2_{200} M_{7.6}\over r_{18} }\right)^{1/2} \;
M_{\odot}/yr, \label{HH}\ee 

\noindent where we have replaced $\pi\Sigma r^2,$ with $\Sigma$ being
the surface density, by $M_D,$ the disk mass interior to radius $r.$
For NGC~4258, we find from (\ref{HH}) and (\ref{MDOT}) that

\begin{equation}
\alpha\sim 10^{-4} \frac{M}{M_D} \left({T^2_{200} M_{7.6}\over r_{18}}
\right)^{-1/2} .
\label{ALPHA} \end{equation}

\noindent The importance of self--gravity is measured by the Toomre
$Q$ parameter, $Q\sim (HM)/ (r M_D),$ such that $Q \ge 1$ is required
for stability to axisymmetric ring modes. From (\ref{THICK}) we find
that the disk becomes gravitationally unstable in this way if $M_D/M
\ge 1.5\times 10^{-3}.$ From (\ref{ALPHA}) we find $\alpha \sim 0.1$
for $M_D/M= 10^{-3}$. A smaller value of $\alpha$ would imply larger
$\Sigma$, small Q value and a more unstable disk.

\noindent Based on the above estimates, we consider a model which is
marginally self--gravitating with $Q \sim 1$ and $\alpha \sim 0.1$
such that $M_D \sim 3 \times 10^4$~$M_\odot$. We note that Maoz~(1995)
has suggested that the radial intervals between successive maser
sources may be related to the wavelength of transient spiral structure
in a marginally self--gravitating disk. The value of $\alpha$ we have
assumed is comparable to that expected to occur in a marginally self
gravitating disk as a result of non--axisymmetric waves (Laughlin~\&
Rozyczka~1996).

\noindent The inner part of the disk of NGC~4258 has been modelled by
Lasota {\it et al.}~(1996) as being advection--dominated. In this case
there would be greater mass flow rate through the disk than indicated
by the observed luminosity. In principle this situation could be
obtained by taking a somewhat thicker more massive disk together with
a larger value of $\alpha.$ As indicated below, our results may be
scaled to apply to such a case.

\noindent Here, we consider the response of a self--gravitating disk,
with parameters like those discussed above, to an orbiting binary
companion in an inclined circular orbit. This interaction produces
warping and precession of the disk. We note that angular momentum
transfer between the companion and disk may produce truncation and gap
formation in the disk (Lin~\& Papaloizou~1993), terminating the radial
progression of maser sources.

\noindent Pringle~(1996) has indicated that forces due to radiation
pressure may be important in producing warps in disks. Such forces
could be included in the formalism presented here. However, we shall
limit ourselves in this application to discussing the effects arising
from a companion only.

\noindent In our model, the companion then needs to have a mass
comparable to that of the disk and an orbital radius somewhat larger
than the size of the disk. Black hole binary systems in which the
companions can have up to comparable mass and separations on a scale
of $10^{18}$~$cm$ have been considered by Begelman, Blandford~\&
Rees~(1980), and postulated to account for jet behavior in
extragalactic sources by Kaastra~\& Roos~(1992) and Roos, Kaastra~\&
Hummel~(1993). The dense cores of globular clusters may be an
alternative candidate for such a companion. From the negative
intrinsic period derivative of millisecond pulsars, Phinney~(1992)
derived a mass $> 4.5 \times 10^3$~$M_\odot$ which is concentrated
within $0.05$~$pc$ from the core of M15. It is possible that such a
core would not only be adequate to provide the necessary perturbation
for the observed warp but also survive the tidal disruption by the
central black hole.

\subsection{Numerical Results}

\noindent For illustrative purposes, we present a numerical model to
show that the observed properties of the warped disk in NGC~4258 can
be regulated by the perturbation on a marginally self--gravitating
disk due to a low--mass companion on an inclined orbit. Since we
include the secular effect of the companion in an orbit with finite
inclination through the perturbing potential $\Psi'_{ext},$ we
consider equation~(\ref{PT17}) with $\omega_z=0$ (i.e. the left hand
side is zero).

\noindent In our numerical calculations, we solve
equation~(\ref{PT23}) by dividing the interval $\left[R_i,R \right]$
into a grid $\left( r_j \right)_{j=1 \ldots n}$ using a constant
spacing $\Delta r$ in the radial direction and such that $r_1=R_i +
\Delta r/2$ and $r_n=R-\Delta r/2$.  We approximate
equation~(\ref{PT17}) at $r=r_j, j=1 \ldots n$ as a system of $n$
linear equations for the $n$ quantities $g_j \equiv g(r_j), j=1 \ldots
n$ (typically $n=200$--2000 in our calculations). The fact that $g_j
=$~constant is a solution gives a solubility condition analagous
to~$(\ref{prec})$ from which the precession frequency may be
found. The perturbing potential can then be modified so as to include
the effects of the Coriolis force. The specification $g_1=0,$
corresponding to zero disk inclination at the inner boundary, then
enables the solution to be completed.

For computational convenience, we normalize the units such that $M=1$,
$R=1$ and $\Omega_K(R)=1.$ The observations of Miyoshi {\it et
al.}~(1995) indicate that $M_D(R)$ of the disk is less than
$10^{-2}M.$ Marginally self--gravitating disks are then such that the
disk semi--thickness is less than one percent of the radius. For
illustrative purposes we have chosen a model for which $\Sigma =
\Sigma_0(R/r -1),$ with the constant $\Sigma_0$ being chosen such that
$M_D(R)=1.5 \times 10^{-3}M, \alpha =0.1,$ and the maximum value of
$H/r$, $(H/r)_{max},$ to be $\sim 10^{-3}$ (see discussion in
\S~\ref{sec:intro}).

The equilibrium structure of the disk is calculated adopting a
polytropic index $1.5$. We plot the Toomre parameter $Q=\kappa c_s/
(\pi G \Sigma),$ with $c_s$ being evaluated in the midplane for our
model disk with $R_i=0.1R$ in Figure~\ref{fig1}. The minimum value of
$Q$ is approximately unity making the disk marginally
self--gravitating.

\placefigure{fig1}

For the companion providing the perturbation, we adopt an orbital
inclination with respect to the disk of $\delta=\pi/4.$ Results for
other values of $\delta$ can be obtained by noting that to a
reasonable approximation, the precession frequency is proportional to
$\cos \delta$ and $g$ is proportional to $\sin 2\delta.$ We also take
the companion's mass to be such that $M_p=M_D(R)$ and the orbital
radius to be $D=1.5R.$

From equation~(\ref{PT17}), we note that $g$ is essentially
independent of a scaling which reduces $M_D, M_p,$ and $H$ by the same
factor. This scaling relation holds because $Q$ is unaltered.
However, for large $\alpha > H/r,$ the small amount of twist induced
by the viscosity is reduced by the same factor. Other things being
equal, $g$ scales with $M_p.$ Results were found to be insensitive to
the form of the surface density taper at the outer boundary.

The induced precession frequency in our model is $\omega_p=-7.9 \times
10^{-5} (J/J_{orb}) \Omega(R).$ We are able to reproduce a warp with a
relatively large amplitude using a perturbing potential which is much
smaller than that due to the central mass, justifying the assumption
of a small perturbation. An approximate estimate for $|g|$ may be
obtained by estimating the magnitudes of terms in~(\ref{ppsip2})
and~(\ref{PT17}) as

\be 
|g|\sim \frac{3M_p R^3}{ 8M_D(R) D^3} |\sin 2 \delta |.
\ee 

\noindent This amplitude can be interpreted as being the product of
the companion induced precession frequency and the time required for a
bending wave to propagate through the disk. When this product is
small, there is good communication between the different parts of the
disk, and the warp is small. In the thin, self--gravitating disks we
have considered here, wave propagation is mainly regulated by
self--gravity which acts to preserve the disk's intrinsic spin vector.

In Figure~\ref{fig2}.a, we plot both the real part of $g$ and the
imaginary part of $-g$ as functions of $r.$ These quantities represent
the inclination of the disk as seen edge on, viewed looking down the
$y$ and $x$--axis, where $\varphi =\pi/2$ and 0 respectively. We have
also calculated $g$ with just the self--gravitating term
in~(\ref{PT18}). As expected (see \S~\ref{sec:viscosity}), the real
part of $g$ so obtained is hardly distinguishable from what we get
when pressure and viscosity with $\alpha=0.1$ are included. For
comparison, Figure~\ref{fig2}.b shows the real part of $g$ obtained
including both self--gravity and pressure terms but with very small
$\alpha =10^{-5}.$ When $\alpha =0$ the solution looks almost exactly
the same. The case $\alpha=0.1$ has been displayed again on the same
plot for comparison. We observe that when $\alpha$ is small, the
response oscillates around that associated with the larger $\alpha.$
This is due to the fact that when both self--gravity and pressure are
present, both a slow and a fast bending wave can propagate in the disk
(Papaloizou~\& Lin~1995). When $\alpha$ is large, the pressure term
in~(\ref{PT18}) is small, so that only the fast wave corresponding to
the purely self--gravitating case can propagate. In Figure~\ref{fig2}.b
we see the superposition of the fast and slow waves, the former having
a larger wavelength than the latter. We have checked that the slow
wave disappears once $\alpha$ becomes larger than $H/r.$ We note that
the precession frequency is the same in all cases. Hereafter, unless
otherwise specified, we shall refer to the case where self--gravity,
pressure and viscosity with $\alpha=0.1$ are included.

\placefigure{fig2}

In Figure~\ref{fig3}, we plot a three dimensional view of the warped
disk. For clarity, the vertical scale has been magnified.
 
\placefigure{fig3}

\noindent The observations indicate the presence of a warp
corresponding to $|g|$ reaching values of about a few tenths at the
disk edge. The results presented here reproduce a warp of the required
magnitude.  Since the outer region of the disk is mainly regulated by
the external perturbation, the functional dependence of $g$ on $r$ in
the outer part of the disk is insensitive to the choice of $R_i$. But
the total range of $|g|$ increases as $R_i$ decreases. This dependence
is due to the fact that {\it in our model} the wave crossing time
across the disk decreases as $R_i$ increases. In the limit of small
$R_i$, it becomes difficult for the outer parts of the disk to
``communicate'' with the inner parts such that the two regions become
essentially disconnected.

\noindent We now consider the twist (the azimuth of the maximum of the
real part of $g$ at a given radius as a function of radius). This
azimuth is given by $2\pi - \eta (r),$ with $\eta (r)$ being the
argument of $g.$ Figure~\ref{fig4} shows $\eta (r)$ as a function of
$r.$

\placefigure{fig4}

\noindent The total variation in $\eta$ is about a few degrees and
most of this variation occurs in the inner regions. A positive
gradient $d \eta (r) / dr$ implies a negative gradient in the azimuth
and a regression of the line of nodes which corresponds to a trailing
twist. This trailing twist is produced by the second term
of~(\ref{PT18}) when viscosity is added (see
\S~\ref{sec:viscosity}). In the outer part of the disk, the magnitude
of this term is a factor $H/r$ smaller than the other terms.  If it is
completely neglected, as noted above, $|g|$ is hardly changed,
indicating that its {\it only} significant contribution is to produce
the small twist. The twist becomes larger in the inner part of the
disk because the perturbation amplitude of the external forcing is
weakened relative to the effect of inertial response and viscous
stress which are contained in the second term of~(\ref{PT18}). The
value of $\eta$ at small $r$ would then be smaller if $Q$ were not as
large there as it is in our model.

\noindent Accurate determination of the twist can provide useful data
on the magnitude of $\alpha$. According to the results in
Figure~\ref{fig4}, the magnitude of $r d \eta /d r \sim 1^{\arcdeg}$
in the outer part of our model disk (with $\alpha = 0.1$) and an order
of magnitude larger in the inner part. In the analysis of NGC~4258,
Herrnstein {\it et al.}~(1996) constructed models to fit the observed
position and velocity of the maser sources.  Some of their models
include variations, along the radial direction, in both the
inclination and the projected position angles of the disk's spin
axis. For example, their model~3 is consistent with a small trailing
twist in which $\vert r d \eta / dr \vert \sim 2^{\arcdeg},$ although
models without any twist fit the data equally well.  We note that the
warp amplitude is on the order of a tenth of the radius throughout
this particular model.

\section{Discussion}
\label{sec:discussion}

We have derived the tilt equation governing the inclination of a near
Keplerian disk subject to global warping $(m=1)$ perturbations taking
into account self--gravity and pressure. We have considered the case
where the perturbation is due to a companion on an inclined orbit, but
the analysis we have presented in this paper is valid for any $m=1$
perturbing potential proportional to the coordinate $z$ along the disk
rotation axis. All effects that can cause evolution on the fastest
possible timescale on the order of $H \Omega r^{-1}$ have been
included.  We also considered the effects of viscosity, showing that
pressure effects diminish in importance relative to those due to
self--gravity once the viscosity parameter $\alpha > H/r.$ The disk
response and precession frequency have been derived using two
different approaches that we have shown to be equivalent in the small
inclination limit. We considered i) the excitation of small amplitude
low--frequency bending modes about the orbital plane of the
companion. Here, the time averaged potential of the companion is
included in the equilibrium state, ii) the zero frequency response of
a disk forced by the secular potential due to a companion in an orbit
with finite inclination. It was found that when the precession
frequency is small compared to the rate of wave propagation across the
disk, the disk can undergo quasi rigid body precession with a small
warp. Then the timescale for the inclination to decrease is found to
be long compared to other timescales in the problem.

For an illustrative application, we have explored the possibility that
the recently observed warped disk in the active galaxy NGC~4258 is
produced by a binary companion. Our results indicate that it can be
produced by a companion with a comparable mass to that of the observed
disk. For a disk mass of $10^{-3}M = 4 \times 10^4$~$M_{\odot}$, the
required mass for the companion is comparable to that contained in the
cores of the densest globular clusters.  Either these cores or black
holes with comparable masses can survive the tidal disruption of the
central black hole.  The warp remains of modest magnitude as long as
the companion induced precession time is everywhere significantly
longer than the time required for a density wave to propagate through
the disk.

If the disk and companion's mass are comparable, their tidal
interaction would lead to the tidal truncation of the outer disk edge
and the companion's orbital evolution on the viscous timescale of the
disk. The exchange of angular momentum between the companion and the
disk occurs through the excitation of density waves in the disk.  For
the parameters we adopted, the viscous evolution timescale of the disk
is $ \sim 8\times 10^8$~$yr$ giving a significant lifetime in the
current state.

Subject to the effect of dynamical friction, the companion may also
undergo orbital decay as it interacts with the field stars in the
nucleus of NGC~4258. For illustrative purposes, we adopt a model in
which the density of the galactic core varies like $r^{-n}$, $r$ being
the distance from the galactic centre. Assuming that the velocity of
the companion stays Keplerian as it spirals down to the centre, the
timescale $t_d$ on which the angular momentum $L$ of the companion
changes is then (Binney~\& Tremaine~1987)

\begin{displaymath}
t_d = \frac{L}{dL/dt} \sim \frac{C}{4 \pi} \frac{1}{\omega}
\frac{M}{M_p} \frac{M}{\rho_0 D^3}
\end{displaymath}

\noindent where $\rho_0$ is the stellar density at the distance $r=D$
from the centre. $C$ is a number which depends on various parameters
of the galactic core (see Binney~\& Tremaine~1987 for more
details). It is typically on the order of unity, and we then set
$C=1$. In our model, $D \sim 0.3$~$pc$. If we take for $\rho_0$ the
value estimated for the galaxy M32, which has been resolved down to
subparsec scale, we have $\rho_0$ between $10^6$ (Lauer {\it et
al.}~1992) and $10^7$~$M_{\odot}/pc^3$. This gives $t_d$ between
$10^7$ and $10^8$~$yr,$ and thus the companion loses angular momentum
because of dynamical friction faster than it gains it by interaction
with the disk. For $t_d$ to be at least on the order of the viscous
timescale of the disk, we need $\rho_0 < 1.5\times
10^5$~$M_{\odot}/pc^3$. This is not unreasonable, since we might
expect the stars in the centre of NGC~4258 to be cleared out by the
presence of the companion.

The fact that the companion spirals in with a lifetime comparable or
smaller than the viscous timescale of the disk provides an additional
argument why we do not expect the disk to have settled to equilibrium
with zero inclination to the orbital plane (see discussion in
\S~\ref{sec:intro}).

\acknowledgments The authors wish to thank L.J. Greenhill for useful
conversations. This work is supported by PPARC through grant
GR/H/09454 and by NASA through grant NAG~53059. C.T. acknowledges
support for this work by the Center for Star Formation Studies at
NASA/Ames Research Center and the University of California at Berkeley
and Santa--Cruz, and by the EU.

\appendix
\section{APPENDIX}

\noindent We here show that the magnitude of the combination of the
two terms in equation~(\ref{PT9}) given by

\be \int^{\infty}_{-\infty} \rho f'_h dz = \int^{\infty}_{-\infty}
\left( \rho \; {\partial \Psi'_{Gh} \over \partial z} + \rho'_h \;
{\partial \Psi_{G} \over \partial z} \right) dz ,
\label{APT1} \ee

\noindent for global perturbations under conditions for which the
Toomre parameter $Q\sim 1,$ is of order $\Omega_K^2 \Sigma H^2 \xi_z
/r^2$ for characteristic values of $H$ and $r$ and so can be
neglected in our ordering scheme. Of course, because these terms
derive from the disk self--gravity, for $Q\gg 1,$ they are even less
significant. We suppose that the disk material is all contained within
$-H< z < H,$ there being vaccuum outside this region. Here we and
below we find it convenient to take $H$ to be the maximum value of the
semi--thickness of the disk.

\noindent We begin by finding $\Psi'_{Gh}.$ We remark that interior to
the disk $\Psi'_{Gh}$ satisfies the Poisson equation

\be \nabla^2 \Psi'_{Gh} = 4 \pi G \rho'_h , \label{APT2} \ee

\noindent while external to the disk it satisfies Laplace's equation.
For the warping perturbations we consider, $\rho'_h$ and $\Psi'_{Gh}$
are odd functions of $z.$ Thus $\partial \Psi'_{Gh} / \partial z$ is
an even function of $z$ and accordingly it has the same values on the
top and bottom of the disk.

\noindent To solve~(\ref{APT2}), we use the standard method of Hankel
transforms (see, e.g., Binney~\& Tremaine~1987). That is we write
(after removing the factors $\exp(i\varphi +\sigma t)$):

\begin{eqnarray}
\Psi'_{Gh} & = & \int^{\infty}_{0}\Psi'_k(k,z) J_m(kr) \; k \; dk,
\label{HT1} \\ \rho'_{h} & = & \int^{\infty}_{0}\rho'_k(k,z) J_m(kr)
\; k \; dk,
\label{HT2} \end{eqnarray}

\noindent where $ \rho'_k(k,z) $ and $\Psi'_k(k,z)$ are the Hankel
transforms of $ \rho'_h(r,z) $ and $\Psi'_{Gh}(r,z)$ respectively. The
Bessel function is denoted by $J_m$ in standard notation. Here $m=1.$
For the global disturbances we are interested in, the significant
values of $k \sim r^{-1}$ at some characteristic radius.

\noindent The Poisson equation~(\ref{APT2}) gives

\be {d^2 \Psi'_{k} \over dz^2} - k^2 \Psi'_{k}= 4\pi G \rho'_k .
\label{APT3} \ee

\noindent Above the disk, where there is no source $(\rho'_k =0),$ the
solution is $\Psi'_k(k,z) = C(k)\exp(-kz),$ where $C(k)$ is an
arbitrary function which depends only on $k.$ Because $\Psi'_k$ is
antisymmetric with respect to reflection in the midplane, below the
disk ($z<0$) we have $\Psi'_k(k,z) =-C(k)\exp(kz).$

\noindent To find $C(k)$, we multiply~(\ref{APT3}) by $z$ and
integrate through the thin disk to obtain

\be 2\left(H{d \Psi'_{k} \over dz}-\Psi'_{k}\right)_+
-k^2\int^{H}_{-H}z\Psi'_{k}dz = 4\pi
G\int^{\infty}_{-\infty}z\rho'_kdz , \label{APT4} \ee

\noindent where the subscript $+$ denotes evaluation in the vaccuum at
the upper boundary of the disk where $z=H.$

\noindent From~(\ref{APT4}) we find to within a multiplicative error
of at most of order $|kH|,$ this being a small quantity for global
perturbations:

\be C(k) = -2\pi G\int^{\infty}_{-\infty}z\rho'_kdz . \label{APT14}
\ee

\noindent From this it similarly follows that at $z=H,$

\be \left({\partial \Psi'_{k} \over \partial z}\right)_{+} = 2\pi G k
\int^{\infty}_{-\infty}z\rho'_kdz . \label{APT5} \ee

\noindent From equation~(\ref{HT1}), it similarly follows that at
$z=H,$

\be \left({\partial \Psi'_{Gh} \over \partial z}\right)_{+}= 2\pi G
\int^{\infty}_{-\infty} \int^{\infty}_{0}z\rho'_k
J_m(kr) \; k^2 \; dk \; dz. \label{HT3}\ee

\noindent We recall that for global density perturbations we expect
that $\rho'_{k}$ is significant only for $k\sim r^{-1}$ at a
characteristic radius.

\noindent Returning to the expression of interest given by
equation~(\ref{APT1}), we first use Poisson's equation to express
densities in terms of potentials:

\be \rho= {\nabla^2\Psi_G \over 4\pi G}, \; \; \; \rho'_{h}=
{\nabla^2\Psi'_{Gh} \over 4\pi G}. \label{APT7} \ee

\noindent However, we shall retain only the derinatives with respect
to $z$ in the Poisson operators which leads to a multiplicative error
of order $H^2/r^2$ for globally varying functions. This in turn leads
to a multiplicative error of order $H/r$ in the final expression we
obtain so it may be neglected. Using~(\ref{APT7}) in this way
in~(\ref{APT1}) we obtain \be \int^{\infty}_{-\infty} \rho f'_h dz =
{1\over 4\pi G}\int^{H}_{-H} \left( {\partial^2 \Psi_{G} \over
\partial z^2}\; {\partial \Psi'_{Gh} \over \partial z} + {\partial^2
\Psi'_{Gh} \over \partial z^2} \; {\partial \Psi_{G} \over \partial z}
\right) dz ={1\over 2\pi G}\left( {\partial \Psi'_{Gh} \over \partial
z} {\partial \Psi_{G} \over \partial z} \right)_{+} ,
\label{APT8} \ee

\noindent where an integration by parts has been performed and the
symmetry properties of the potentials have been used.

\noindent We also note that for the equilibrium potential 

\be \left({\partial \Psi_{G} \over \partial z} \right)_{+} =2\pi
G\Sigma .  \ee

\noindent Using this together with~(\ref{HT3}) in~(\ref{APT8}), we
obtain

\be \int^{\infty}_{-\infty} \rho f'_h dz = 2\pi G\Sigma
\int^{\infty}_{-\infty} \int^{\infty}_{0}z\rho'_k
J_m(kr) \; k^2 \; dk \; dz. \label{HT4}\ee

\noindent Using (\ref{HT2}) this may be equivalently expressed as

\be \int^{\infty}_{-\infty} \rho f'_h dz = 2\pi
G\Sigma\int^{\infty}_{-\infty} z\rho'_h dz
\left[{\int^{\infty}_{-\infty} \int^{\infty}_{0}z\rho'_k J_m(kr) \;
k^2 \; dk \; dz \over \int^{\infty}_{-\infty}
\int^{\infty}_{0}z\rho'_k J_m(kr) \; k \; dk \; dz}\right].
\label{HT5}\ee

\noindent We remark that the expression in square brackets gives a
mean value of $k$ which for global perturbations may be taken $\sim
r^{-1}$ at some characteristic radius.  Further from the expression
(\ref{PT13}) we can simply estimate that for global perturbations \be
\int^{\infty}_{-\infty}z\rho'_hdz \sim {\Sigma H \xi_z\over r} .  \ee

\noindent These taken together with the condition $Q\sim 1,$ yields
the required estimate

\be \int^{\infty}_{-\infty} \rho f'_h dz \sim \Omega_K^2 \Sigma H^2
\xi_z /r^2. \ee

\newpage

\begin{figure}
\plotone{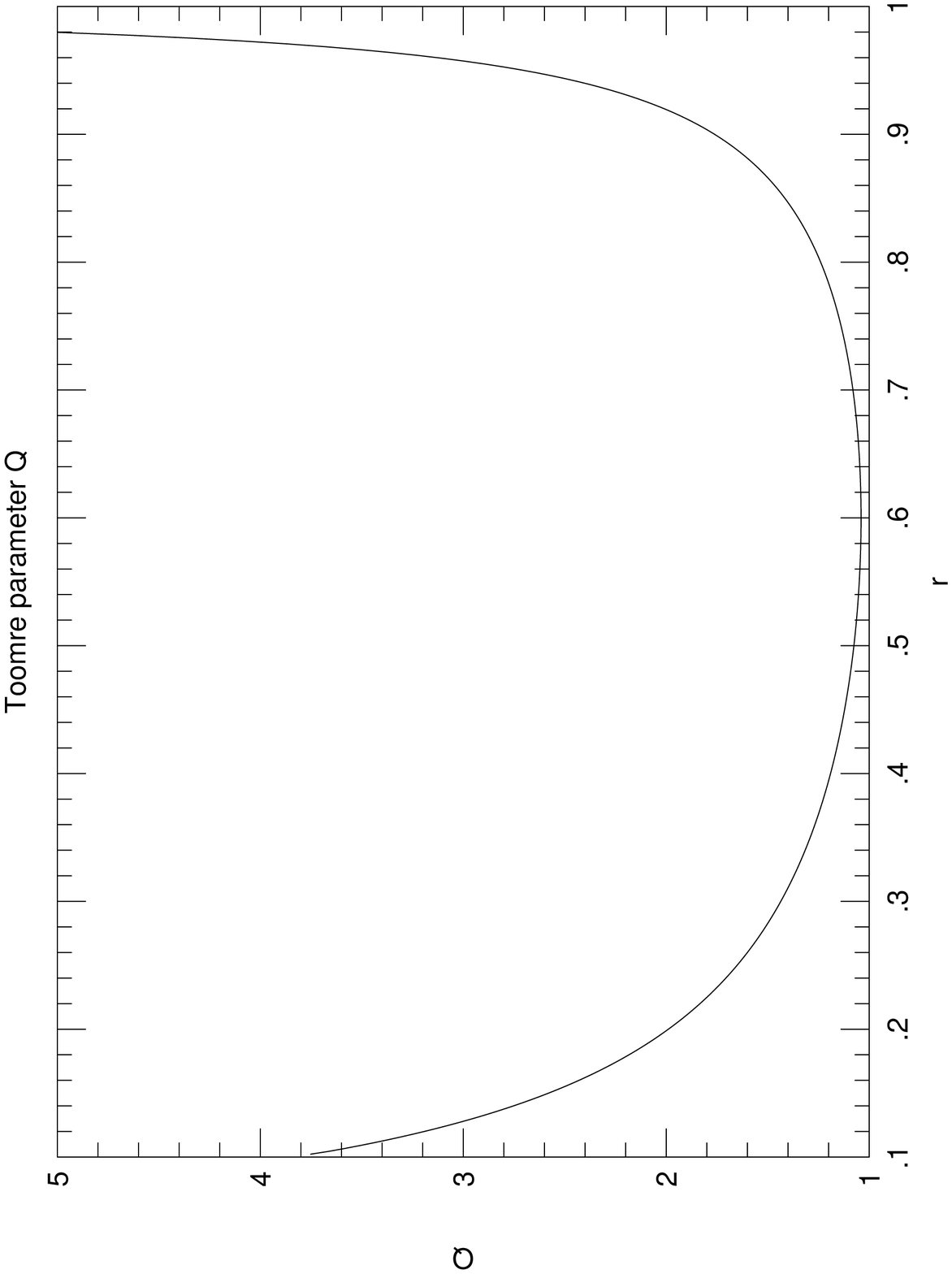}
\caption[]{Toomre parameter $Q=\kappa c/ (\pi G \Sigma)$ for
$M_D(R)=1.5.10^{-3}M$ and $(H/r)_{max}=1.5.10^{-3}$. The other
parameters are given in the text.}
\label{fig1}
\end{figure}

\begin{figure}
\plotone{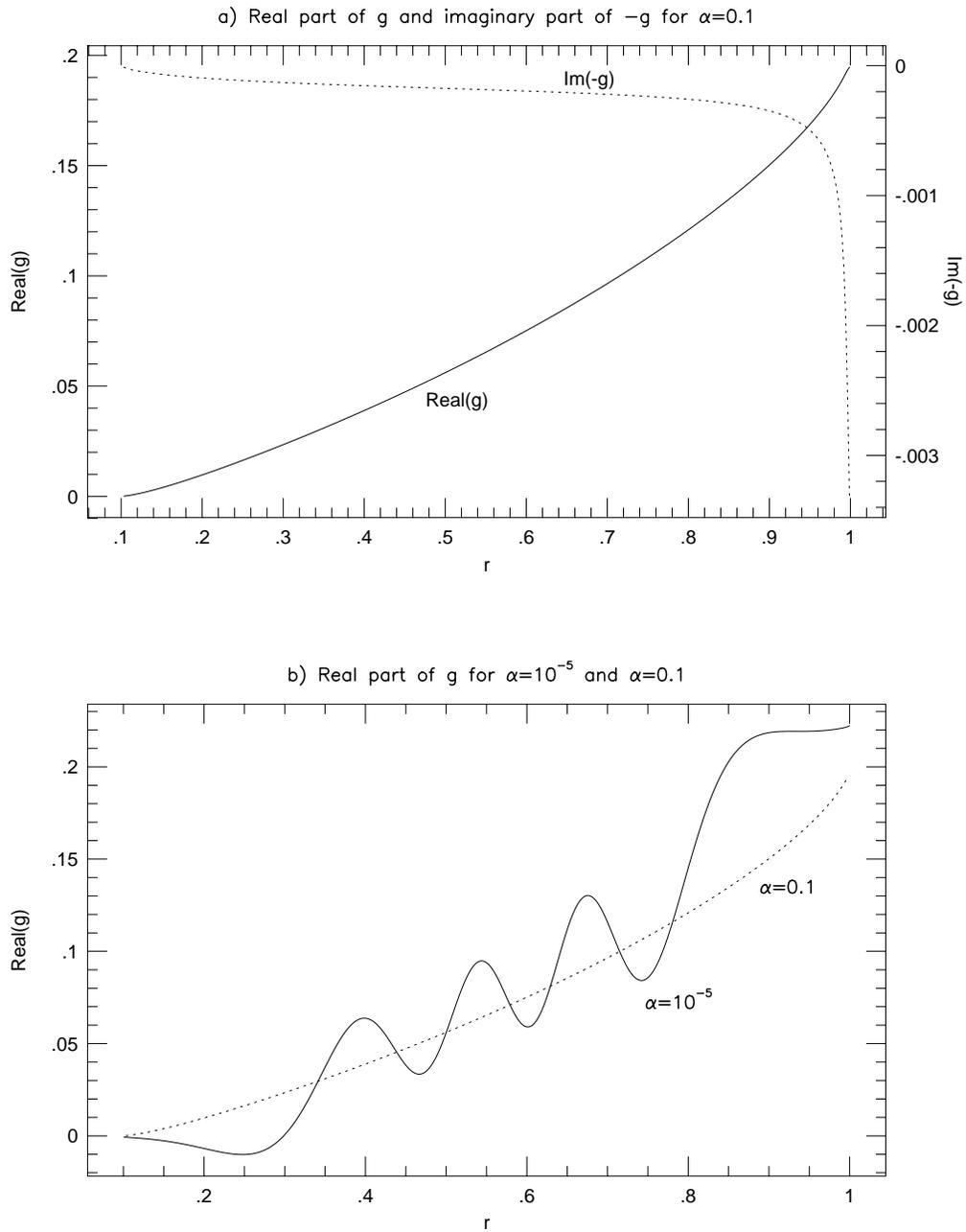}
\caption[]{a) Real and imaginary parts of $g$ and $-g$ respectively for a
self--gravitating viscous disk with $\alpha=0.1$. The other parameters
are given in the text. b) Real part of $g$ for the same disk but
$\alpha=10^{-5}$ and $\alpha=0.1$.}
\label{fig2}
\end{figure}

\newpage
\topmargin -1.cm

\begin{figure}
\plotone{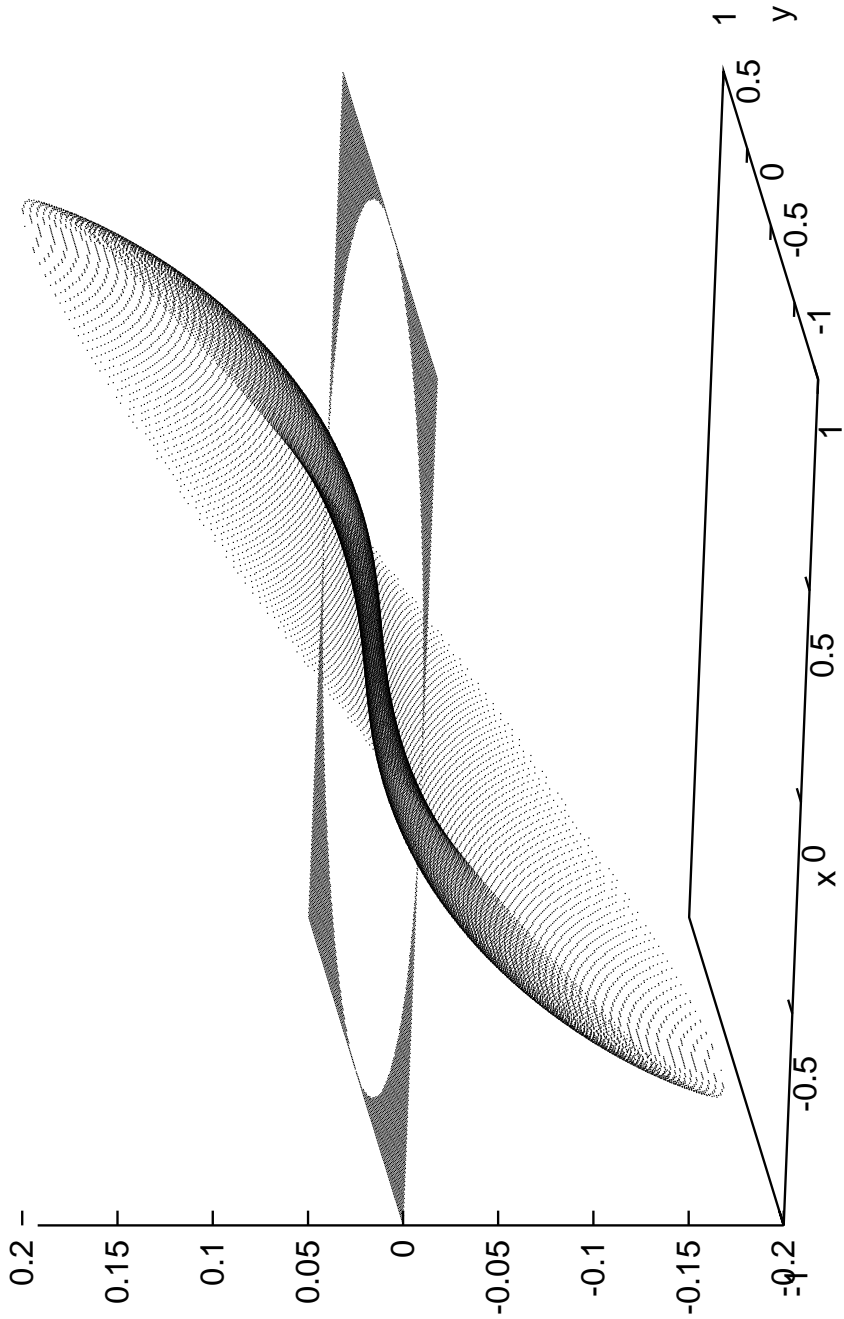}
\caption[]{3D view of the warped disk, with parameters given in the
text. The vertical scale has been magnified. The plane $z=0$
represents the plane of the undistorted disk.}
\label{fig3}
\end{figure}

\newpage

\begin{figure}
\plotone{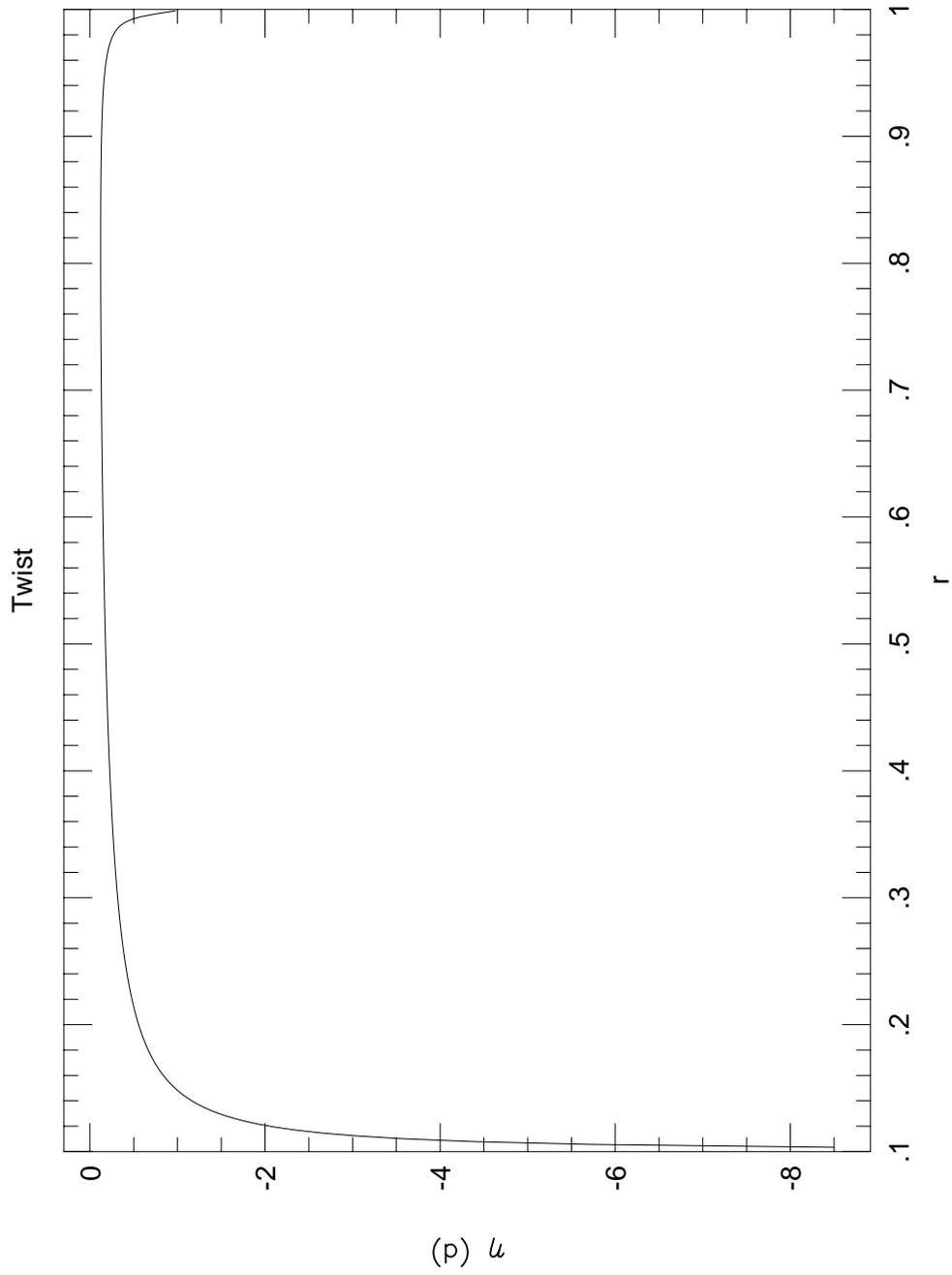}
\caption[] {The phase angle $\eta(r),$ the variation of which defines
the degree of twist, plotted in degrees as a function of $r.$}
\label{fig4}
\end{figure}

\end{document}